\documentclass[structabstract,a4paper]{aa}
\usepackage{graphicx}
\usepackage{natbib}
\usepackage{color}
\usepackage{lscape}
\usepackage{txfonts}

\def\arcm{\hbox{$^\prime$}}
\def\arcs{\hbox{$^{\prime\prime}$}}
\def\farcs{\hbox{$.\!\!^{\prime\prime}$}}
\def\fm{\hbox{$.\!\!^{\rm m}$}}
\def\degr{\hbox{$^\circ$}}
\def\fdg{\hbox{$.\!\!^\circ$}}

\def\aj{AJ}
\def\apj{ApJ}
\def\apjs{ApJS}
\def\mnras{MNRAS}
\def\pasp{PASP}
\def\aa{A\&A}

\bibpunct{(}{)}{;}{a}{}{,}

\begin{document}

\title{Galactic bulge giants: probing stellar and galactic evolution}
\subtitle{I. Catalogue of {\em Spitzer} IRAC and MIPS sources\thanks{Table~5
as well as the fits mosaics are only available in electronic form at the CDS
via anonymous ftp to cdsarc.u-strasbg.fr (130.79.128.5) or via
http://cdsweb.u-strasbg.fr/cgi-bin/qcat?J/A+A/}}
\author{
  Stefan Uttenthaler\inst{1}
  \and
  Matthias Stute\inst{2,3,4}
  \and
  Raghvendra Sahai\inst{2}
  \and
  Joris A.\ D.\ L.\ Blommaert\inst{1}
  \and
  Mathias Schultheis\inst{5}
  \and
  Kathleen E.\ Kraemer\inst{6}
  \and
  Martin A.\ T.\ Groenewegen\inst{7}
  \and
  Stephan D.\ Price\inst{6}
}

\institute{
  Instituut voor Sterrenkunde, K.\ U.\ Leuven, Celestijnenlaan 200D, 3001
  Leuven, Belgium\\
  \email{stefan@ster.kuleuven.be}
  \and
  Jet Propulsion Laboratory, California Institute of Technology, 4800 Oak
  Grove Drive, Pasadena, CA 91109, USA
  \and
  IASA and Section of Astrophysics, Astronomy and Mechanics, Department of
  Physics, University of Athens, Panepistimiopolis, 157 84 Zografos, Athens,
  Greece
  \and
  Dipartimento di Fisica Generale "A.\ Avogadro", Universita degli Studi di
  Torino, Via Pietro Giuria 1, 10125 Torino, Italy
  \and
  Observatoire de Besan\c{c}on, 41bis, avenue de l'Observatoire, 25000
  Besan\c{c}on, France
  \and
  Air Force Research Laboratory, Space Vehicles Directorate, 29 Randolph Rd.,
  Hanscom AFB, MA 01731, USA
  \and
  Royal Observatory of Belgium, Ringlaan 3, 1180 Brussels, Belgium
}

\date{Received March 31, 2009; accepted January 15, 2010}

\abstract
{}
{We aim at measuring mass-loss rates and the luminosities of a statistically
  large sample of Galactic bulge stars at several galactocentric radii.
  The sensitivity of previous infrared surveys of the bulge has been rather
  limited, thus fundamental questions for late stellar evolution, such as the
  stage at which substantial mass-loss begins on the red giant branch and its
  dependence on fundamental stellar properties, remain unanswered. We aim at
  providing evidence and answers to these questions.}
{To this end, we observed seven $15\times15$\,arcmin$^2$ fields in the nuclear
  bulge and its vicinity with unprecedented sensitivity using the IRAC and MIPS
  imaging instruments on-board the Spitzer Space Telescope. In each of the
  fields, tens of thousands of point sources were detected.}
{In the first paper based on this data set, we present the observations, data
  reduction, the final catalogue of sources, and a detailed comparison to
  previous mid-IR surveys of the Galactic bulge, as well as to theoretical
  isochrones. We find in general good agreement with other surveys and the
  isochrones, supporting the high quality of our catalogue.}
{}

\keywords{Galaxy: bulge --- Galaxy: stellar content --- Infrared: stars ---
Stars: late-type --- Stars: mass-loss --- Stars: AGB and post-AGB}

\maketitle


\section{Introduction}

The Galactic bulge (GB), an important dynamical and morphological component of
our Galaxy, offers an environment distinct from the Galactic disk for study of
stellar populations, stellar evolution, and the mass-loss processes that
accompany and, in the end, control the last. Understanding and calibrating the
physical processes whereby mass ejected by evolved stars into the bulge
environment is recycled back into new generations of stars requires a
statistical knowledge of mass loss as a function of fundamental stellar
parameters in this region. Because of the limited sensitivity of previous
surveys of the bulge, fundamental questions for late stellar evolution, such as
the stage at which substantial mass-loss begins on the red giant branch (RGB),
and its dependence on fundamental stellar properties, remain unanswered. The GB
is an ideal laboratory for addressing these issues, providing a very large
sample of stars at an almost identical distance.

We therefore observed seven $15\times15$\,arcmin$^2$ fields that sample a
range of distances from the Galactic centre with unprecedented sensitivity
using the Infrared Array Camera \citep[IRAC;][]{Faz04} and the Multiband Imaging
Photometer for {\em Spitzer} \citep[MIPS;][]{Rie04}, the imaging instruments
on-board the {\em Spitzer} Space Telescope \citep{Wer04}, in order to determine
mass-loss rates and luminosities of a statistically large sample of stars at
several galactocentric radii. These data enable us to detect stars with very
low mass-loss rates through their infrared excess, determine the dependence of
the mass-loss rate on luminosity and effective temperature along the giant
branches, and conduct a census of mass-losing stars at different rates. The
observations, together with existing studies that probe higher mass-loss rate
stars, will enable us to infer the total rate of mass loss in the bulge, a key
input to evolutionary models of the bulge. The data have already led to the
discovery of mid-IR $\log P$ vs.\ magnitude relations \citep{Gla09}. In this
paper we present the observations, the data reduction, and the source
catalogue. We also compare the data to previous mid-IR surveys of the bulge and
to theoretical isochrones.

The outline of this paper is as follows. In \S\ref{sec_obs}, we describe the
observations, the data reduction, the different steps of point source
extraction, and how the catalogues were created. We then proceed in
\S\ref{sec_surveys} with checking our photometric data against other mid-IR
surveys of the Galactic bulge. Further checks are presented in \S\ref{sec_cmds},
where colour-magnitude diagrams (CMDs) are compared to theoretical isochrones.
Finally, \S\ref{sec_conclu} draws conclusions on the catalogues and the data
quality.

\section{Observations and Data Reduction} \label{sec_obs}

\subsection{Field selection and observations}

The locations of the observed fields were chosen to sample the bulge on a
variety of scales, to measure how the mix of stellar populations varies with
Galactic latitude. They were also chosen to avoid the relative intense,
saturating emission from near the Galactic plane. Our innermost fields are
within the central stellar cusp, which presumably contains stars of various
ages \citep[c.f.][]{Blu03,Fig04}. Because the stars within this domain are
believed to have formed within the central molecular zone and then to have
diffused into an increasingly thicker distribution as a result of scattering
off molecular clouds \citep{Kim01}, we expect a vertical segregation of stellar
ages. Thus, inasmuch as the luminosities and mass-loss rates of red giants and
asymptotic giant branch (AGB) stars depend on their masses, hence their ages,
the radial distributions of the different kinds of evolved stellar objects can
be used to model the star formation and dynamical history of this region.

\begin{figure}
\includegraphics[width=\columnwidth]{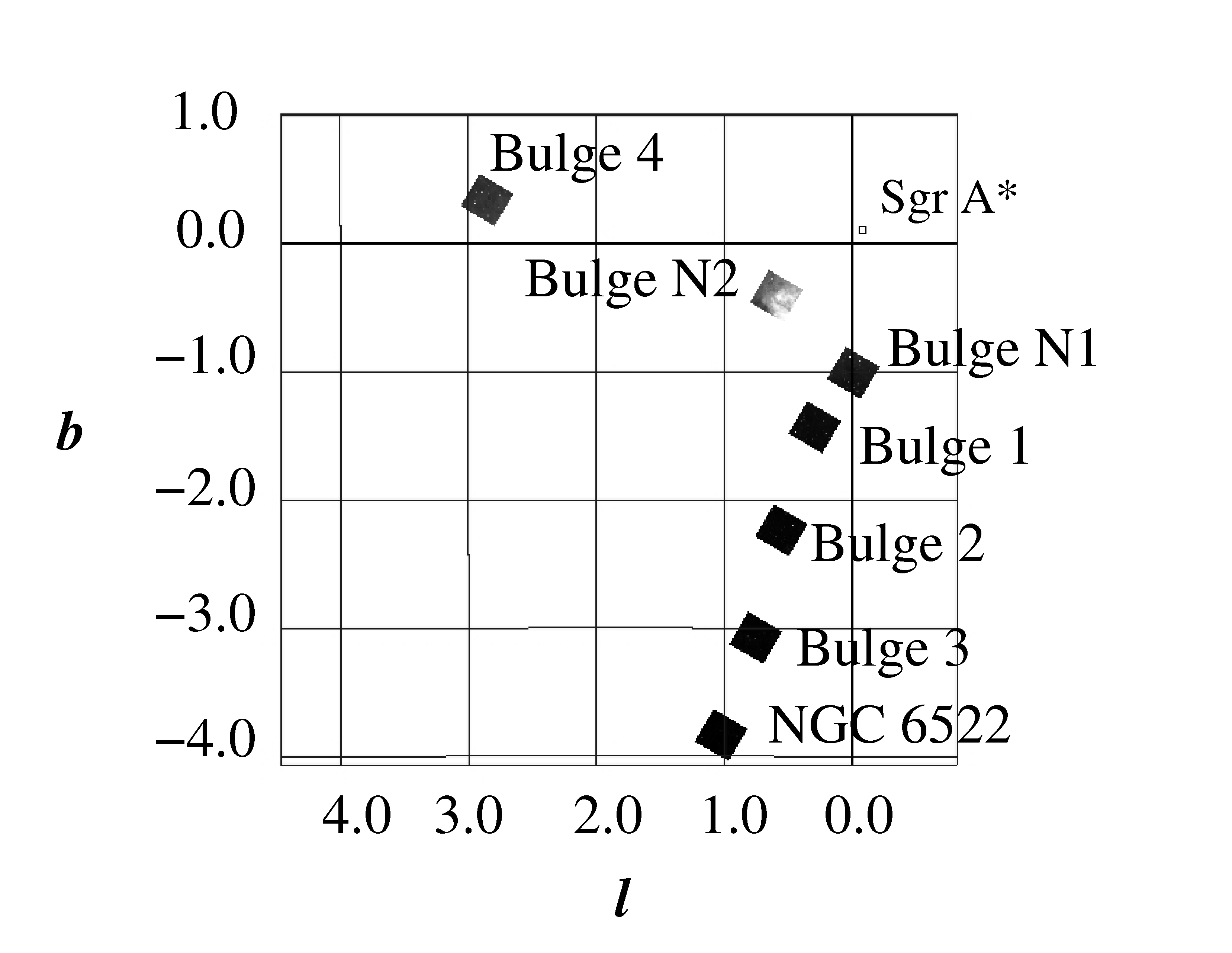}
  \caption{Map of the seven fields observed in this campaign with the labels
    used throughout this paper.}
  \label{fields}
\end{figure}

Two fields were selected to sample the nuclear bulge at
$(l,b)=(0\fdg00,-1\fdg00)$ and $(l,b)=(0\fdg63,-0\fdg36)$, namely N\,1 and
N\,2, both of which were observed by ISOGAL \citep[project for imaging part of
the Galaxy using ISO, the Infrared Space Observatory;][]{Omo03} at 7 and
15\,$\mu$m. Five fields were selected beyond the nuclear bulge in areas where
they overlap the Optical Gravitational Lensing Experiment III
\citep[OGLE-III;][]{Uda00} micro-lensing survey. Four of these are located below
the Galactic plane along a radial vector that subtends the minor axis of the
Galaxy at an angle of about 13\fdg5, and terminates at the well-studied field in
Baade's window at $(l,b)=(1\fdg03,-3\fdg83)$, containing the globular cluster
\object{NGC\,6522} \citep[e.g.][]{Gla99}. The three fields inside this are
located at $(l,b)=(0\fdg30,-1\fdg42)$, $(0\fdg56,-2\fdg23)$, and
$(0\fdg76,-3\fdg07)$. The fifth field is positioned above the plane at
$(l,b)=(2\fdg87,0\fdg35)$. All fields are approximately rectangular in right
ascension (RA) and declination (Dec) coordinates. Figure~\ref{fields} shows the
location of the seven observed fields with respect to the Galactic centre, and
Table~\ref{tbl_fields} summarises some of their main characteristics. The range
in RA and Dec (J2000) given in Table~\ref{tbl_fields} refers to where there is
full overlap between MIPS and all four IRAC bands. Particularly in IRAC we do
have some coverage outside the given range.

\begin{table*}
\caption{Main characteristics of the seven fields towards the GB observed with
  {\em Spitzer} as part of programme 2345.}
\label{tbl_fields}
\centering
\begin{tabular}{lcccrcc}
\hline\hline
Field name & RA range (h m s) & Dec range ($\degr\ \arcm\ \arcs$) & l centre & b centre & Date IRAC obs & Date MIPS obs \\
\hline
Bulge 1    & 17 51 14.2 \dots 17 52 36.7 &  $-$29 15 22 \dots $-$29 34 15 & 0.30 & $-$1.42 & 2005 03 30 & 2005 04 09 \\
Bulge 2    & 17 55 02.1 \dots 17 56 24.2 &  $-$29 26 25 \dots $-$29 45 13 & 0.56 & $-$2.23 & 2005 03 31 & 2005 04 10 \\
Bulge 3    & 17 58 50.7 \dots 18 00 13.3 &  $-$29 41 08 \dots $-$29 59 59 & 0.76 & $-$3.07 & 2005 03 30 & 2005 04 10 \\
Bulge 4    & 17 50 16.1 \dots 17 51 36.4 &  $-$26 08 56 \dots $-$26 27 51 & 2.87 &    0.35 & 2005 03 30 & 2005 04 09 \\
Bulge N\,1 & 17 48 51.9 \dots 17 50 14.8 &  $-$29 17 42 \dots $-$29 36 42 & 0.00 & $-$1.00 & 2005 03 30 & 2005 04 08 \\
Bulge N\,2 & 17 47 55.8 \dots 17 49 16.3 &  $-$28 28 41 \dots $-$28 46 19 & 0.63 & $-$0.36 & $-$        & 2005 04 13 \\
NGC\,6522  & 18 02 30.4 \dots 18 03 53.5 &  $-$29 49 26 \dots $-$30 08 24 & 1.03 & $-$3.83 & 2005 03 30 & 2005 04 10 \\
\hline
\end{tabular}
\end{table*}

Observations of the GB fields were performed using the IRAC instrument in all
four channels at 3.6, 4.5, 5.8, and 8.0\,$\mu$m, and the MIPS instrument in the
24\,$\mu$m channel on board the {\em Spitzer} Space Telescope within the
programme 2345. Bulge field N\,2 was observed only with MIPS in our programme,
because this field was covered by IRAC observations within the general observer
(GO) programme 3677 \citep[principal investigator: S.\ Stolovy;][]{Ram08}. The
IRAC observations were carried out on 2005 March 30 and 31, and the MIPS
observations between 2005 April 08 and 13.

With IRAC, observations were done in the full-array read out-mode, one frame
per pointing with 2\,s integration time per frame -- the shortest full-array
integration time was used to minimise the effects of saturation and resulting
latency problems. The mapping was done in a $6\times6$ rectangular grid with
step size 260\arcs, with five dither positions, and a medium scale factor,
giving a total exposure time of 10\,s per pixel. The MIPS observations
were obtained using the photometry/raster mode with 10\,s integration time and
full-array read-out mode, $3\times3$ rectangular grid and two cycles,
giving a total integration time of 331\,s per pixel.

The fields of view of IRAC channels 1 and 3, 2 and 4, as well as MIPS did not
fully overlap. Nevertheless, the MIPS field of view is fully contained in all
four IRAC fields of view.

\subsection{Data reduction}

We corrected the IRAC basic calibrated data (bcd) files to mitigate artifacts
such as muxbleed or column pulldown with the tools provided by S.\ Carey at the
{\em Spitzer} Science Center (SSC). Post-bcd processing was then conducted on
the corrected bcd files using the MOsaicker and Point source EXtractor (MOPEX)
software and its subsystem APEX \citep[version 18.2.0,][]{MaM05}.

\subsubsection{Step 1: Point source detection and extraction}

Before further processing, mosaics were created by MOPEX for each field and
channel from the corrected bcd frames. In this step, the pipeline interpolates
the input images onto the output grid, taking geometric distortion into account.
An outlier detection scheme flags bad pixels and any pixels affected by cosmic
ray hits or moving objects, and these pixels are re-computed before co-addition.
Finally, the interpolated images are co-added to one mosaic image. After
mosaicking, APEX determines the background by calculating the median in a
$45\times45$ pixel box around each pixel, and subtracts it from the image.
These background-subtracted images were used in the detection step. Then,
background fluctuations in the images were estimated and noise images derived,
which were used for signal-to-noise ratio (SNR) estimation and the generation
of point source probability images. With those, the detection table was
compiled. We chose a detection threshold of $2.3 \sigma$ above the background.
We used both the point response function (PRF) fitting capability and the
aperture photometry functionality with circular apertures with radii of 2, 3,
and 4 pixels. Larger apertures were not applicable owing to the crowding in the
fields. The tiles used for PRF fitting also had to be chosen to be very small
($3\times3$ pixels). We used the most up-to-date mean PRFs as provided by the
SSC, and set the PRF normalisation radius accordingly.

The measured $\chi^2$ values in our data seem to be very high, even for 
successfully fitted sources, e.g.\ in the least-crowded field NGC\,6522, they
are between 2.5 and 3.3 in the IRAC channels 1 and 2, between 1.4 and 1.8 in
IRAC\,3, and only in IRAC\,4 at the perfect value of 1. This may stem from
confusion noise, which is not included in the provided uncertainty images
and/or from the high source density. Confusion noise is created by the
amplitude variations from PRFs of closely spaced sources
\citep{Hac87,Rie95}. Modelling the confusion noise has not improved the
resulting $\chi^2$ values. This has consequences for the reliability of
automatic de-blending procedures, which were not working properly in this
field. We thus had to disable the active and passive de-blending capabilities
of APEX (see Appendix~\ref{sec_blending}).

In all MIPS fields, we first detected bright sources, removed the Airy rings of
these sources to do the source detection on the residual images, and performed
the point source extraction on the original images. The southern half of the
field N\,2 has strong diffuse emission. In that area, the detection threshold
is certainly higher than in areas with low diffuse background emission, as is
also evident from the histograms of magnitudes in Appendix~\ref{sec_histo}.

\subsubsection{Step 2: Assigning quality grades to the sources}
\label{sec_qualgr}

\begin{figure}[!ht]
\includegraphics[width=8.7cm]{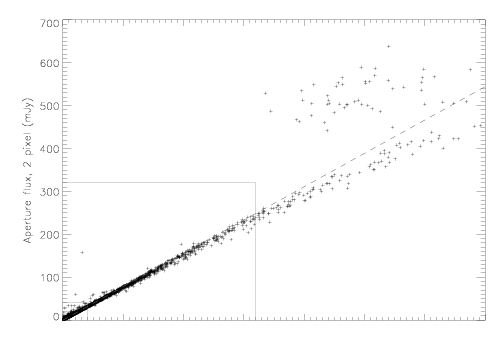}\\
\includegraphics[width=8.7cm]{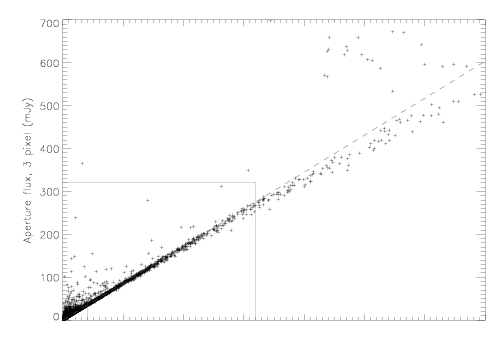}\\
\includegraphics[width=8.7cm]{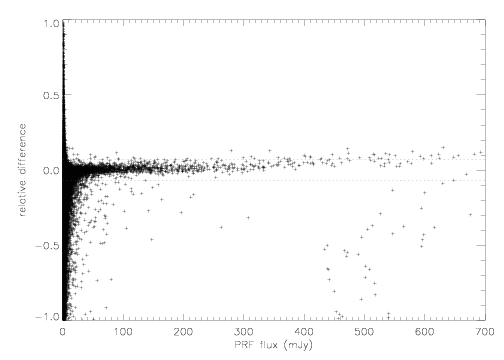}\\
\caption{Comparison of the fluxes resulting from PRF fitting and aperture 
  photometry for the IRAC\,1 observations of the NGC\,6522 field. Top:
  PRF fitting fluxes and aperture photometry for the aperture with a radius of
  2 pixels; middle: the same as top panel, but for the aperture with a radius
  of 3 pixels; bottom: relative difference between the measured fluxes (3
  pixels aperture radius) versus the PRF flux for all sources. The dotted lines
  in the bottom panel mark relative differences of $\pm7$\%. In the upper and
  middle panels, the dashed line is a linear least squares fit to the sources
  with $\chi^{2} < 5$, and the solid horizontal and vertical lines mark the
  lower and saturation limits of Table~\ref{tbl_limits} used for quality grade
  labelling.}
\label{errors}
\end{figure}

Three different effects can be seen when plotting the flux obtained by PRF
fitting  versus that found with aperture photometry for all the extracted
sources (Fig.~\ref{errors}). The first effect is present when we use a large
aperture with a radius of 3 or 4 pixels. At very low PRF fluxes, there are
several sources that show disproportionately large aperture fluxes, i.e.\ are
above the dashed line. These sources are faint sources close to a bright
source. In case of the larger aperture, we therefore already get a
significant contribution in the aperture from the bright source. The second
effect is relevant for some bright sources, whose PRF fluxes are smaller than
the aperture fluxes. These sources are super-saturated; i.e., the radial
intensity profile has a dip in the centre and so it shows a double peak
structure. In a few cases, depending on the depth of the central valley, these
two peaks are identified as two individual sources that are each fitted with a
fainter PRF; i.e., both sources lie above the dashed line. In other cases, the
fitted PRF ends up with a central flux between that of the central valley and
that of the wings, meaning PRF fitting underestimates the total flux, so these
sources are also shifted to the left in Fig.~\ref{errors}.

The SSC homepage provides values of the maximum flux of unsaturated point
sources as a function of {\em integration} time. Table~\ref{tbl_limits}
presents the flux values interpolated to a frame {\em exposure} time of 1.2\,s,
along with a lower flux limit. This lower flux limit was adopted on subjective
grounds for the quality grade labelling (see below). We thus have to be
cautious with sources brighter than about 320\,mJy in IRAC\,1. Consistent with
this warning is a third slight effect that seems to set in at fluxes close to
this value and affects all brighter sources: the PRF fluxes increase
disproportionately compared to the aperture fluxes, leading to a bending to the
right, away from the dashed line in Fig.~\ref{errors} (top). Aperture
photometry thus measures the saturated plateau of the PRF, while PRF fitting
seems to ignore the missing flux that would be present beyond the plateau and
fits the wings. The individual source then lies below the dashed line, with a
larger PRF flux than the aperture flux.

\begin{table}
\caption{Flux limits applied for assigning quality grades.}
\label{tbl_limits}
\centering
\begin{tabular}{ccc}
\hline\hline
Band & Lower limit (mJy) & Saturation limit (mJy) \\
\hline
IRAC 3.6\,$\mu$m & 40 & 320  \\
IRAC 4.5\,$\mu$m & 30 & 330  \\
IRAC 5.8\,$\mu$m & 20 & 2300 \\
IRAC 8.0\,$\mu$m & 20 & 1200 \\
MIPS  24\,$\mu$m & 40 & 220  \\
\hline
\end{tabular}
\end{table}

If dividing the aperture flux by the linear least squares fit to sources with
$\chi^{2} < 5$, the fluxes measured with both ways are almost identical for
many of the sources with $\chi^{2} < 5$. The slope of this fit is not
the aperture correction factor, to be derived in Sect.~\ref{sec_apt_corr}.
Except for the sources affected by the three aforementioned effects, the
relative difference calculated as
\begin{equation}
{\rm rel.\ difference} = \frac{{\rm f}_{\rm PRF} - {\rm f}_{\rm aperture}^{\rm 
corrected}}{{\rm f}_{\rm PRF}}
\label{equ_relerr}
\end{equation}
scatters around 7\% (Fig.~\ref{errors}, bottom). The superscript ``corrected''
in Eq.~\ref{equ_relerr} indicates that the aperture flux has been divided by
the linear fit to sources with $\chi^{2} < 5$. Sources that are within this
range are considered as quality grade~A. This test is performed first,
therefore grade~A sources span the whole range of fluxes, from faint to almost
saturated sources.

Faint sources with fluxes smaller than a certain lower limit
(Table~\ref{tbl_limits}) and outside the 7\% range are classified as quality
grade~B. Sources whose fluxes are above the saturation limit are classified as
quality grade~C. They would have to be treated differently with dedicated tools.
However, since these sources are probably mostly foreground stars, they are not
of prime interest to the science goals, thus we decided not to invest more time
in their flux determination. The interested reader is invited to download and
analyse the mosaics, which will be made available as on-line material at Centre
de Donn\'{e}es astronomiques de Strasbourg (CDS).

Sources that failed all of the above criteria are classified as quality
grade~D. These are blended sources for which the PRFs overlap such that even
the  aperture with the smallest radius includes both peaks; i.e., no reliable
flux  measurement is available for these sources. This effect is the same as
the first one mentioned above related to the larger aperture with a radius of 3
pixels or larger.

The statistics of grades for all seven fields and all five bands are given in 
Table~\ref{tbl_stats_grades}. The percentage given for grades A and B is
relative to the total number of sources in the catalogue for the respective
field that is given in the last column. The seventh column gives the
number of sources not detected in the respective channel, but detected in at
least one of the other bands plus in the reference catalogues (see
Sect.~\ref{sect_catalogue}).

\begin{table*}
\caption{Statistics of sources with their grades for all seven fields and
five bands.}
\label{tbl_stats_grades}
\centering
\begin{tabular}{ccrrrrrr}
\hline\hline
Field & Band & Grade A & Grade B & Grade C & Grade D & Not detected  & total \\
\hline
Bulge 1  & 3.6\,$\mu$m & 12446 (18\%) & 39177 (57\%) &   183 &   626 & 16661 & 69093 \\
         & 4.5\,$\mu$m & 16173 (23\%) & 35598 (52\%) &    85 &   216 & 17021 &       \\
         & 5.8\,$\mu$m & 13839 (20\%) & 21757 (31\%) &     1 &   169 & 33327 &       \\
         & 8.0\,$\mu$m & 12669 (18\%) & 17044 (25\%) &     7 &    96 & 39277 &       \\
         & 24\,$\mu$m  &   577 ( 1\%) &  1197 ( 2\%) &     1 &     2 & 67316 &       \\
Bulge 2  & 3.6\,$\mu$m & 14889 (25\%) & 30525 (51\%) &   197 &   748 & 13963 & 60322 \\
         & 4.5\,$\mu$m & 17495 (29\%) & 28190 (47\%) &    72 &   492 & 14073 &       \\
         & 5.8\,$\mu$m & 13712 (23\%) & 17900 (30\%) &     0 &   214 & 28496 &       \\
         & 8.0\,$\mu$m & 11960 (20\%) & 15752 (26\%) &     1 &   152 & 32457 &       \\
         & 24\,$\mu$m  &   503 ( 1\%) &  1119 ( 2\%) &     4 &     4 & 58692 &       \\
Bulge 3  & 3.6\,$\mu$m & 13476 (26\%) & 27927 (53\%) &   131 &   101 & 11031 & 52666 \\
         & 4.5\,$\mu$m & 16329 (31\%) & 25159 (48\%) &    62 &    64 & 11052 &       \\
         & 5.8\,$\mu$m & 12085 (23\%) & 15880 (30\%) &     0 &    35 & 24666 &       \\
         & 8.0\,$\mu$m & 10712 (20\%) & 13220 (25\%) &     4 &    24 & 28706 &       \\
         & 24\,$\mu$m  &   383 ( 1\%) &   976 ( 2\%) &     1 &     2 & 51304 &       \\
Bulge 4  & 3.6\,$\mu$m & 12938 (22\%) & 33046 (57\%) &   136 &   197 & 12091 & 58408 \\
         & 4.5\,$\mu$m & 15689 (27\%) & 29961 (51\%) &    75 &    95 & 12588 &       \\
         & 5.8\,$\mu$m & 11311 (19\%) & 18283 (31\%) &     3 &   177 & 28634 &       \\
         & 8.0\,$\mu$m &  9239 (16\%) & 13588 (23\%) &    19 &    98 & 35464 &       \\
         & 24\,$\mu$m  &   267 ( 0\%) &   817 ( 1\%) &     3 &     4 & 57317 &       \\
Bulge N\,1 & 3.6\,$\mu$m & 12553 (19\%) & 35826 (55\%) &   198 &   724 & 15534 & 64835 \\
         & 4.5\,$\mu$m & 15402 (24\%) & 34527 (53\%) &    85 &   296 & 14525 &       \\
         & 5.8\,$\mu$m & 11819 (18\%) & 21845 (34\%) &     5 &   520 & 30646 &       \\
         & 8.0\,$\mu$m & 12035 (19\%) & 16565 (26\%) &    11 &   144 & 36080 &       \\
         & 24\,$\mu$m  &   467 ( 1\%) &  1126 ( 2\%) &     3 &    23 & 63216 &       \\
Bulge N\,2 & 24\,$\mu$m  &    21 ( 5\%) &   316 (74\%) &    17 &    74 &     0 &   428 \\
NGC\,6522 & 3.6\,$\mu$m & 12430 (26\%) & 24924 (53\%) &    79 &    55 &  9796 & 47284 \\
         & 4.5\,$\mu$m & 15637 (33\%) & 21263 (45\%) &    32 &    21 & 10331 &       \\
         & 5.8\,$\mu$m & 12011 (25\%) & 11907 (25\%) &     0 &    30 & 23336 &       \\
         & 8.0\,$\mu$m & 10263 (22\%) & 10432 (22\%) &     0 &    12 & 26577 &       \\
         & 24\,$\mu$m  &   431 ( 1\%) &  1152 ( 2\%) &     0 &     0 & 45701 &       \\
\hline
\end{tabular}
\end{table*}

\subsubsection{Step 3: Determination of the aperture corrections}
\label{sec_apt_corr}

Since we do not use a larger aperture to estimate the background, which would 
give unreliable photometry in the crowded fields, but do aperture photometry on
the background-subtracted images instead, the aperture corrections are expected
to be  different from the standard values given in the instrument handbooks. 

To derive aperture corrections, we created point spread functions with
stinytim2.0 \citep{Kri06} for all five bands.
We then inserted artificial sources with known flux in our mosaicked,
background-subtracted images and compared the resulting fluxes given by APEX
with the initial fluxes with which we had created the sources. The slope of a
linear least squares fit to the initial vs.\ measured flux gives the aperture
correction factors. The derived aperture corrections for the aperture with two
pixels radius
are given in Table~\ref{tbl_aper_corr}, along with their statistical error as
derived from the linear fit. Also given in the table are the values as
recommended by the SSC, for a 2 pixel radius on source and a 10 -- 20 pixel
radius background annulus in the case of IRAC, and a 3\farcs5 on source
aperture without any background annulus in the case of MIPS. The SSC
recommended values are not directly comparable to our aperture correction
factors, so they are only given for illustrative purposes. Despite the high
value of the aperture correction factor for MIPS\,24, we adopted the small
aperture with 2 pixels radius because i) our observations are not photon noise
limited, thus a larger aperture would not reduce the error estimate; ii) there
is a ``population'' of blended sources whose flux increases disproportionally
with growing aperture because more of the neighbouring source's flux is
measured with larger aperture; iii) the SSC lists aperture corrections for an
even smaller aperture, so our value is not an extreme. We also corrected
the PRF fluxes with the photometric correction factors as recommended by the
SSC.

\begin{table}
\caption{Aperture correction factors}
\label{tbl_aper_corr}
\centering
\begin{tabular}{ccc}
\hline\hline
Band & Aperture correction & SSC recommended \\
\hline
IRAC 3.6\,$\mu$m &  1.2129$\pm$0.0008 & 1.205 \\
IRAC 4.5\,$\mu$m &  1.2730$\pm$0.0005 & 1.221 \\
IRAC 5.8\,$\mu$m &  1.4693$\pm$0.0007 & 1.363 \\
IRAC 8.0\,$\mu$m &  1.6556$\pm$0.0006 & 1.571 \\
MIPS  24\,$\mu$m &  1.5742$\pm$0.0007 & 2.560 \\
\hline
\end{tabular}
\end{table}

Besides the aperture correction and the photometric correction for the PRF
fluxes, we did not apply any other correction factors such as the colour
correction. According to the IRAC data handbook, the colour correction factor
for blackbody spectra with temperatures between 2\,000 and 5\,000\,K is around
1\%, except for IRAC\,4, where it can reach as much as 2.7\% for the hotter
stars. Since the colour correction is small and we do not know a priori the
nature and spectrum of our objects, we did not apply these corrections.

\subsubsection{Step 4: Creating the point source catalogue}
\label{sect_catalogue}

We finally decided to adopt only the PRF fluxes for the final catalogue
on the basis of comparison with other catalogues and with isochrones in CMDs.
With this choice, we find, on the one hand, slightly better agreement with
other catalogues, and on the other, somewhat reduced scatter at the faint end
of CMDs. See Sect.~\ref{sec_surveys} and \ref{sec_cmds} for a comparison
with other catalogues and isochrones.

The last step is the cross-identification, for each field, of the extraction
tables of all five bands among each other and with other reference catalogues.
As reference catalogues we adopt the Deep Near Infrared Survey
\citep[DENIS;][]{Epc97}, the 2~Micron All Sky Survey \citep[2MASS;][]{Skr06},
the Midcourse Space Experiment (MSX) catalogue, and the Infra-Red Astronomical
Satellite (IRAS) point source catalogue. We applied the following criteria to
include a source in the final catalogue. We identified individual sources by
their RA and Dec position and allowed for an error margin of 1\farcs6, i.e.\
slightly more than one pixel size, for the cross-identification among our five
bands. Sources that are detected in at least two of our five bands are included
in any case in the final catalogue. A cross-identification is made for these
sources with a search radius of 3\farcs0 with DENIS and 2MASS, and 30\farcs0 for
identification with an MSX or IRAS counterpart. The information about these
counterparts is included in the catalogue. Sources that are detected in only one
of our five bands are only included in the final catalogue if they also have a
counterpart in at least one of the reference catalogues, with the same search
radii applied as for the sources with two or more {\em Spitzer} detections.
This procedure was followed to include as many of the sources in the final
catalogue that were only detected in MIPS\,24. These might very well be real
objects, albeit extremely red (i.e.\ not detected with IRAC). Additionally, by
following this procedure, more of the sources that are located in the areas that
are not sampled by all five bands will be included in the catalogue. These areas
are also included in the final catalogue. The least reliable sources in our
catalogue are those with detection in only one IRAC band and no MIPS detection,
and with (i) either no 2MASS and DENIS counterpart (depending on the field,
between 3.1\% and 6.7\% of the sources), or (ii) a DENIS and 2MASS counterpart
at a distance between 1.6 and 3 arcseconds (depending on the field, between
0.4\% and 1.0\% of the sources).

Table~\ref{cattable_1} gives ten lines of the band-merged catalogue of the Bulge
N\,1 field as an example. The entire table will be made available for download
from CDS. A portion is shown here for guidance regarding its form and content.

The columns of our point source catalogue are explained as follows.

\begin{description}

\item[Column 1:] Source identification, IAU-conform identifier.

\item[Column 2:] Right ascension in degrees (J2000).

\item[Column 3:] Declination in degrees (J2000).

\item[Column 4:] PRF flux in IRAC\,1 (in $\mu$Jy), set to ``$-$9.99999e+99'' if
  not detected.

\item[Column 5:] PRF flux in IRAC\,2 (in $\mu$Jy), set to ``$-$9.99999e+99'' if
  not detected.

\item[Column 6:] PRF flux in IRAC\,3 (in $\mu$Jy), set to ``$-$9.99999e+99'' if
  not detected.

\item[Column 7:] PRF flux in IRAC\,4 (in $\mu$Jy), set to ``$-$9.99999e+99'' if
  not detected.

\item[Column 8:] PRF flux in MIPS\,24 (in $\mu$Jy), set to ``$-$9.99999e+99''
  if not detected.

\item[Column 9:] Quality flags  of the IRAC\,1, 2, 3, 4, and MIPS\,24
  channels (see Section~\ref{sec_qualgr}), set to ``$-$'' if not detected.

\item[Column 10:] Uncertainty of the PRF flux in IRAC\,1 (in $\mu$Jy), set to
  ``$-$9.99999e+99'' if not detected.

\item[Column 11:] Uncertainty of the PRF flux in IRAC\,2 (in $\mu$Jy), set to
  ``$-$9.99999e+99'' if not detected.

\item[Column 12:] Uncertainty of the PRF flux in IRAC\,3 (in $\mu$Jy), set to
  ``$-$9.99999e+99'' if not detected.

\item[Column 13:] Uncertainty of the PRF flux in IRAC\,4 (in $\mu$Jy), set to
  ``$-$9.99999e+99'' if not detected.

\item[Column 14:] Uncertainty of the PRF flux in MIPS\,24 (in $\mu$Jy), set to
  ``$-$9.99999e+99'' if not detected.

\item[Column 15:] Observation flag. The first four bits state if the position
  of the source was in the field of view of the IRAC bands (``1'') or not
  (``0''), the fifth bit is for MIPS\,24.

\item[--]

\item[Column 16:] Distance to closest DENIS source (arcsec; only
  \mbox{$<3\farcs0$}), set to ``$-$'' if no counterpart.

\item[Column 17:] DENIS $I$-band magnitude, set to ``$-$'' if no counterpart
  or the counterpart has no $I$-band measurement.

\item[Column 18:] DENIS $J$-band magnitude, set to ``$-$'' if no counterpart
  or the counterpart has no $J$-band measurement.

\item[Column 19:] DENIS $K$-band magnitude, set to ``$-$'' if no counterpart
  or the counterpart has no $K$-band measurement.

\item[--]

\item[Column 20:] Distance to closest 2MASS source (arcsec; only
  \mbox{$<3\farcs0$}), set to ``$-$'' if no counterpart. Only sources with
  quality flags A, B, C, or D in at least one 2MASS filter have been considered.

\item[Column 21:] 2MASS $J$-band magnitude, set to ``$-$'' if no counterpart
  within \mbox{$3\farcs0$} was found, or ``$-9.999$'' if the counterpart has no
  measured $J$-band magnitude available or too low quality flag.

\item[Column 22:] 2MASS $H$-band magnitude, set to ``$-$'' if no counterpart
  within \mbox{$3\farcs0$} was found, or ``$-9.999$'' if the counterpart has no
  measured $H$-band magnitude available or too low quality flag.

\item[Column 23:] 2MASS $K$-band magnitude, set to ``$-$'' if no counterpart
  within \mbox{$3\farcs0$} was found, or ``$-9.999$'' if the counterpart has no
  measured $K$-band magnitude available or too low quality flag.

\item[Column 24:] 2MASS quality flags, set to ``$-$'' if no counterpart or too
  low quality flag.

\item[--]

\item[Column 25:] Distance to closest IRAS source (arcsec; only
  \mbox{$<30\arcs$}), set to ``$-$'' if no counterpart. Only sources with
  quality flags 2 or 3 in at least one IRAS band have been considered.

\item[Column 26:] IRAS 12\,$\mu$m flux (Jy), set to ``$-$'' if no counterpart,
  and to ``$-$9.9e+99'' if the counterpart has no measured flux in the
  12\,$\mu$m band or too low quality flag.

\item[Column 27:] IRAS 12\,$\mu$m quality flag, set to ``$-$'' if no
  counterpart or too low quality flag.

\item[Column 28:] IRAS 25\,$\mu$m flux (Jy), set to ``$-$'' if no counterpart,
  and to ``$-$9.9e+99'' if the counterpart has no measured flux in the
  25\,$\mu$m band or too low quality flag.

\item[Column 29:] IRAS 25\,$\mu$m quality flag, set to ``$-$'' if no
  counterpart or too low quality flag.

\item[Column 30:] IRAS 60\,$\mu$m flux (Jy), set to ``$-$'' if no counterpart,
  and to ``$-$9.9e+99'' if the counterpart has no measured flux in the
  60\,$\mu$m band or too low quality flag.

\item[Column 31:] IRAS 60\,$\mu$m quality flag, set to ``$-$'' if no
  counterpart or too low quality flag.

\item[Column 32:] IRAS 100\,$\mu$m flux (Jy), set to ``$-$'' if no counterpart,
  and to ``$-$9.9e+99'' if the counterpart has no measured flux in the
  100\,$\mu$m band or too low quality flag.

\item[Column 33:] IRAS 100\,$\mu$m quality flag, set to ``$-$'' if no
  counterpart or too low quality flag.

\item[--]

\item[Column 34:] Distance to closest MSX source (arcsec; only
  \mbox{$<30\arcs$}), set to ``$-$'' if no counterpart. Only sources with
  quality flags 2, 3, or 4 in at least one MSX band have been considered.

\item[Column 35:] MSX B1 band flux (Jy), set to ``$-$'' if no counterpart,
  and to ``$-$9.999e+99'' if the counterpart has no measured flux in the
  B1 band or too low B1 band quality flag.

\item[Column 36:] MSX B1 band quality flag, set to ``$-$'' if no counterpart or
  too low quality flag.

\item[Column 37:] MSX B2 band flux (Jy), set to ``$-$'' if no counterpart,
  and to ``$-$9.999e+99'' if the counterpart has no measured flux in the
  B2 band or too low B2 band quality flag.

\item[Column 38:] MSX B2 band quality flag, set to ``$-$'' if no counterpart or
  too low quality flag.

\item[Column 39:] MSX A band flux (Jy), set to ``$-$'' if no counterpart,
  and to ``$-$9.999e+99'' if the counterpart has no measured flux in the
  A band or too low A band quality flag.

\item[Column 40:] MSX A band quality flag, set to ``$-$'' if no counterpart or
  too low quality flag.

\item[Column 41:] MSX C band flux (Jy), set to ``$-$'' if no counterpart,
  and to ``$-$9.999e+99'' if the counterpart has no measured flux in the
  C band or too low C band quality flag.

\item[Column 42:] MSX C band  quality flag, set to ``$-$'' if no counterpart or
  too low quality flag.

\item[Column 43:] MSX D band flux (Jy), set to ``$-$'' if no counterpart,
  and to ``$-$9.999e+99'' if the counterpart has no measured flux in the
  D band or too low D band quality flag.

\item[Column 44:] MSX D band quality flag, set to ``$-$'' if no counterpart or
  too low quality flag.

\item[Column 45:] MSX E band flux (Jy), set to ``$-$'' if no counterpart,
  and to ``$-$9.999e+99'' if the counterpart has no measured flux in the
  E band or too low E band quality flag.

\item[Column 46:] MSX E band quality flag, set to ``$-$'' if no counterpart or
  too low quality flag.

\end{description}


\begin{landscape}
\centering
\begin{table}
\setcounter{table}{4}
\caption{Sample table of the band-merged catalogue of the N\,1 field.
}
\label{cattable_1}
\begin{tabular}{cccccccccc}
\hline\hline
 Source id.\        & RA         & Dec          & IRAC\,1      & IRAC\,2         & IRAC\,3         & IRAC\,4         & MIPS\,24        & Quality flags     & $\Delta$\,IRAC\,1 \\
 (1)                & (2)        & (3)          & (4)          & (5)             & (6)             & (7)             & (8)             & (9)               & (10)              \\
\hline
USB267.50681$-$29.54985 & 267.506807 & $-$29.549854 & 2.707150e+05 &   1.996047e+05 &   2.018591e+05 & $-$9.99999e+99 &   1.293678e+05 & A  A   A  $-$  D  & 6.895201e+02 \\
USB267.50882$-$29.54757 & 267.508822 & $-$29.547572 & 2.885406e+03 &   2.078063e+03 & $-$9.99999e+99 & $-$9.99999e+99 & $-$9.99999e+99 & B  B  $-$ $-$ $-$ & 5.886386e+01 \\
USB267.50743$-$29.55153 & 267.507432 & $-$29.551530 & 4.349657e+03 & $-$9.99999e+99 & $-$9.99999e+99 & $-$9.99999e+99 & $-$9.99999e+99 & B $-$ $-$ $-$ $-$ & 5.367287e+01 \\
USB267.50761$-$29.55523 & 267.507612 & $-$29.555234 & 8.086190e+04 & $-$9.99999e+99 & $-$9.99999e+99 & $-$9.99999e+99 &   2.392784e+06 & D $-$ $-$ $-$  C  & 2.820764e+02 \\
USB267.50862$-$29.55417 & 267.508623 & $-$29.554167 & 1.133203e+05 &   7.665020e+04 &   8.380626e+04 & $-$9.99999e+99 & $-$9.99999e+99 & D  D   D  $-$ $-$ & 3.408423e+02 \\
USB267.50730$-$29.55596 & 267.507301 & $-$29.555959 & 2.159647e+05 &   1.786561e+05 & $-$9.99999e+99 & $-$9.99999e+99 & $-$9.99999e+99 & C  C  $-$ $-$ $-$ & 9.627816e+02 \\
USB267.50860$-$29.55727 & 267.508600 & $-$29.557267 & 3.002938e+05 & $-$9.99999e+99 & $-$9.99999e+99 & $-$9.99999e+99 & $-$9.99999e+99 & C $-$ $-$ $-$ $-$ & 1.635651e+03 \\
USB267.50811$-$29.55797 & 267.508113 & $-$29.557972 & 1.587659e+05 & $-$9.99999e+99 & $-$9.99999e+99 & $-$9.99999e+99 & $-$9.99999e+99 & C $-$ $-$ $-$ $-$ & 9.000979e+02 \\
USB267.50573$-$29.55593 & 267.505733 & $-$29.555935 & 8.429971e+04 &   6.759881e+04 &   1.763209e+05 & $-$9.99999e+99 & $-$9.99999e+99 & D  D   D  $-$ $-$ & 4.045054e+02 \\
USB267.50479$-$29.55561 & 267.504790 & $-$29.555611 & 4.142997e+04 & $-$9.99999e+99 & $-$9.99999e+99 & $-$9.99999e+99 & $-$9.99999e+99 & D $-$ $-$ $-$ $-$ & 1.880509e+02 \\
\hline
\end{tabular}
\end{table}

\centering
\begin{table}
\setcounter{table}{4}
\caption{Sample table of the band-merged catalogue of the N\,1 field
  (continued).}
\label{cattable_2}
\begin{tabular}{ccccccccccccccccc}
\hline\hline
$\Delta$\,IRAC\,2 &  $\Delta$\,IRAC\,3 &  $\Delta$\,IRAC\,4 & $\Delta$\,MIPS\,24 & obs.\ flag & & & & & & & & & & & & \\
 (11)           & (12)            & (13)            & (14)            & (15)      & & (16)  & (17)   & (18)   & (19) &  & (20) & (21)     & (22)     & (23)     & (24) & \\
\hline
  5.543478e+02 &   5.596869e+02 & $-$9.99999e+99 &   5.804075e+01 & 11111 & $|$ & 0.431 & 14.684 & 10.946 &  8.699 & $|$ & 0.316 &   10.799 &  9.300 &    8.452 & AAA & $|$ \\
  4.209486e+01 & $-$9.99999e+99 & $-$9.99999e+99 & $-$9.99999e+99 & 11111 & $|$ &   $-$ &    $-$ &    $-$ &    $-$ & $|$ & 0.152 & $-$9.999 & 12.902 &   12.604 & UAA & $|$ \\
$-$9.99999e+99 & $-$9.99999e+99 & $-$9.99999e+99 & $-$9.99999e+99 & 11111 & $|$ & 2.682 & 14.974 &    $-$ &    $-$ & $|$ & 0.370 & $-$9.999 & 13.389 & $-$9.999 & UBU & $|$ \\
$-$9.99999e+99 & $-$9.99999e+99 & $-$9.99999e+99 &   3.545098e+02 & 11111 & $|$ &   $-$ &    $-$ &    $-$ &    $-$ & $|$ &  $-$  &      $-$ &    $-$ &      $-$ & $-$ & $|$ \\
  2.687747e+02 &   3.786693e+02 & $-$9.99999e+99 & $-$9.99999e+99 & 11111 & $|$ & 0.288 & 14.974 & 11.759 &  9.230 & $|$ & 0.281 &   11.225 &  9.678 &    9.084 & AAA & $|$ \\
  6.126482e+02 & $-$9.99999e+99 & $-$9.99999e+99 & $-$9.99999e+99 & 11111 & $|$ & 0.700 & 14.535 &    $-$ &    $-$ & $|$ &   $-$ &      $-$ &    $-$ &      $-$ & $-$ & $|$ \\
$-$9.99999e+99 & $-$9.99999e+99 & $-$9.99999e+99 & $-$9.99999e+99 & 11111 & $|$ & 2.840 &    $-$ &  9.955 &  5.166 & $|$ & 2.781 & $-$9.999 & 13.012 & $-$9.999 & UAU & $|$ \\
$-$9.99999e+99 & $-$9.99999e+99 & $-$9.99999e+99 & $-$9.99999e+99 & 11111 & $|$ &   $-$ &    $-$ &    $-$ &    $-$ & $|$ &   $-$ &      $-$ &    $-$ &      $-$ & $-$ & $|$ \\
  2.687747e+02 &   6.536204e+02 & $-$9.99999e+99 & $-$9.99999e+99 & 11111 & $|$ & 2.717 &    $-$ & 14.089 & 12.045 & $|$ & 2.779 &   13.725 & 12.248 &   11.631 & AAB & $|$ \\
$-$9.99999e+99 & $-$9.99999e+99 & $-$9.99999e+99 & $-$9.99999e+99 & 11111 & $|$ & 0.900 &    $-$ & 14.089 & 12.045 & $|$ & 0.708 &   13.725 & 12.248 &   11.631 & AAB & $|$ \\
\hline
\end{tabular}
\end{table}

\centering
\begin{table}
\setcounter{table}{4}
\caption{Sample table of the band-merged catalogue of the N\,1 field
  (continued).}
\label{cattable_3}
\begin{tabular}{cccccccccccccccccc}
\hline\hline
      &          &      &          &      &            &      &            &      &     &      &           &      &              &      &           &      &      \\
 (25) & (26)     & (27) & (28)     & (29) & (30)       & (31) & (32)       & (33) &     & (34) & (35)      & (36) & (37)         & (38) & (39)      & (40) & (41) \\
\hline
 23.1 & 1.89e+01 &    3 & 1.59e+01 &    3 & $-$9.9e+99 &    1 & $-$9.9e+99 &    1 & $|$ & 26.5 & 9.063e+00 &    3 & $-$9.999e+99 &    0 & 1.678e+01 &    4 & 2.109e+01 \\
  $-$ &      $-$ &  $-$ &      $-$ &  $-$ &        $-$ &  $-$ &        $-$ &  $-$ & $|$ &  $-$ &       $-$ &  $-$ &          $-$ &  $-$ &       $-$ &  $-$ &       $-$ \\
 17.2 & 1.89e+01 &    3 & 1.59e+01 &    3 & $-$9.9e+99 &    1 & $-$9.9e+99 &    1 & $|$ & 20.4 & 9.063e+00 &    3 & $-$9.999e+99 &    0 & 1.678e+01 &    4 & 2.109e+01 \\
  4.6 & 1.89e+01 &    3 & 1.59e+01 &    3 & $-$9.9e+99 &    1 & $-$9.9e+99 &    1 & $|$ &  7.1 & 9.063e+00 &    3 & $-$9.999e+99 &    0 & 1.678e+01 &    4 & 2.109e+01 \\
  9.6 & 1.89e+01 &    3 & 1.59e+01 &    3 & $-$9.9e+99 &    1 & $-$9.9e+99 &    1 & $|$ & 11.6 & 9.063e+00 &    3 & $-$9.999e+99 &    0 & 1.678e+01 &    4 & 2.109e+01 \\
  2.1 & 1.89e+01 &    3 & 1.59e+01 &    3 & $-$9.9e+99 &    1 & $-$9.9e+99 &    1 & $|$ &  4.5 & 9.063e+00 &    3 & $-$9.999e+99 &    0 & 1.678e+01 &    4 & 2.109e+01 \\
  6.9 & 1.89e+01 &    3 & 1.59e+01 &    3 & $-$9.9e+99 &    1 & $-$9.9e+99 &    1 & $|$ &  3.8 & 9.063e+00 &    3 & $-$9.999e+99 &    0 & 1.678e+01 &    4 & 2.109e+01 \\
  7.5 & 1.89e+01 &    3 & 1.59e+01 &    3 & $-$9.9e+99 &    1 & $-$9.9e+99 &    1 & $|$ &  3.6 & 9.063e+00 &    3 & $-$9.999e+99 &    0 & 1.678e+01 &    4 & 2.109e+01 \\
  3.4 & 1.89e+01 &    3 & 1.59e+01 &    3 & $-$9.9e+99 &    1 & $-$9.9e+99 &    1 & $|$ &  6.9 & 9.063e+00 &    3 & $-$9.999e+99 &    0 & 1.678e+01 &    4 & 2.109e+01 \\
  6.5 & 1.89e+01 &    3 & 1.59e+01 &    3 & $-$9.9e+99 &    1 & $-$9.9e+99 &    1 & $|$ & 10.0 & 9.063e+00 &    3 & $-$9.999e+99 &    0 & 1.678e+01 &    4 & 2.109e+01 \\
\hline
\end{tabular}
\end{table}
\clearpage

\centering
\begin{table}
\setcounter{table}{4}
\caption{Sample table of the band-merged catalogue of the N\,1 field
  (continued).}
\label{cattable_4}
\begin{tabular}{ccccc}
\hline\hline
     &           &      &           &      \\
(42) &      (43) & (44) &      (45) & (46) \\
\hline
   4 & 1.698e+01 &    4 & 1.593e+01 &   4 \\
 $-$ &       $-$ &  $-$ &      $-$  & $-$ \\
  4  & 1.698e+01 &    4 & 1.593e+01 &   4 \\
  4  & 1.698e+01 &    4 & 1.593e+01 &   4 \\
  4  & 1.698e+01 &    4 & 1.593e+01 &   4 \\
  4  & 1.698e+01 &    4 & 1.593e+01 &   4 \\
  4  & 1.698e+01 &    4 & 1.593e+01 &   4 \\
  4  & 1.698e+01 &    4 & 1.593e+01 &   4 \\
  4  & 1.698e+01 &    4 & 1.593e+01 &   4 \\
  4  & 1.698e+01 &    4 & 1.593e+01 &   4 \\
\hline
\end{tabular}
\end{table}
\end{landscape}
\clearpage

\section{Comparison with other missions and catalogues}\label{sec_surveys}

Our observations overlap with a number of other mid-IR surveys. These are

\begin{enumerate}

\item ISOGAL: Survey with the ISO Camera (ISOCAM) on-board the ISO satellite,
  combined with DENIS $IJK_S$ photometry.

\item Galactic Legacy Infrared Mid-Plane Survey Extraordinaire II (GLIMPSE-II),
  a {\em Spitzer}/IRAC survey of the area $\pm 1\degr$ around the Galactic
  plane and the Galactic centre \citep{GLIMPSE}.

\item GALactic CENtre (GALCEN), a {\em Spitzer}/IRAC survey of the inner
  $\sim 1\fdg4 \times 2\fdg0$ of the Galaxy \citep{Ram08}.

\item A {\em Spitzer}/MIPS survey of the inner $\sim 1\fdg5 \times 8\fdg0$ of
  the Galaxy \citep{Hin08}.

\end{enumerate}

\subsection{Comparison with ISOGAL}\label{comp_isogal}

ISOGAL \citep{Omo03} is a survey with the ISOCAM instrument on-board the ISO
satellite in two bands at 7 and 15\,$\mu$m, combined with $IJK_{\rm{S}}$
photometry from the DENIS project \citep{Epc97}. The 7\,$\mu$m LW2 band of
ISOCAM ($\sim 5.0 - 8.5$\,$\mu$m, central wavelength 6.75\,$\mu$m) overlaps in
wavelength with the IRAC\,4 channel \citep[$\sim 6.4 - 9.4$\,$\mu$m, nominal
wavelength 7.844\,$\mu$m;][]{Hor08}. For IRAC\,4, we have overlap with ISOGAL
in the fields Bulge\,2, Bulge\,4, N\,1, and NGC\,6522. The number of sources in
common (with less than $2\farcs0$ positional offset) is 163 for Bulge\,2, 335
for Bulge\,4, 288 for N\,1, and 264 in NGC\,6522. We present here a brief
comparison for the NGC\,6522 field, because the ISO 7\,$\mu$m photometry goes
deepest in that field (least crowding and diffuse background emission). The
results for the other fields are very similar.

\begin{figure}
\includegraphics[width=\columnwidth,bb=70 368 540 556]{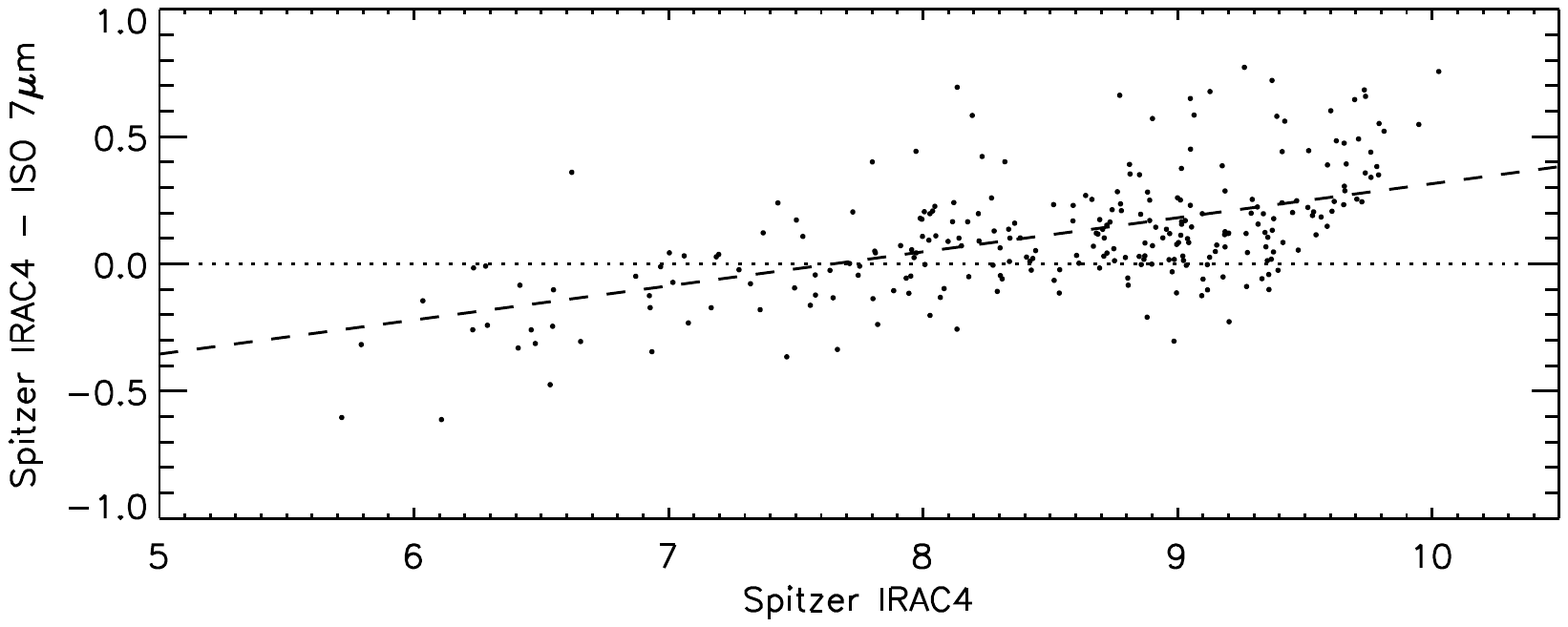}
\caption{Magnitude difference between the {\em Spitzer} IRAC\,4 band and the
  ISO 7\,$\mu$m LW2 band for 264 sources in common in the NGC\,6522 field. The
  dotted line marks the identity relation, the dashed line is a linear least
  squares fit through the data points, excluding those sources fainter than
  9\fm3 in either of the data sets to avoid Malmquist bias.}
\label{ISOGAL_NGC6522}
\end{figure}

Figure~\ref{ISOGAL_NGC6522} shows the magnitude difference between the
{\em Spitzer} IRAC\,4 band and the ISO LW2 band for the NGC\,6522 field. The
data points do not scatter randomly around zero. Rather, faint sources tend to
be brighter in the ISO 7\,$\mu$m band, whereas bright sources tend to be
brighter in the IRAC\,4 band. The relation found to convert between the
two bands is
\begin{equation}
mag_{\rm{ISO}\,7} = 0.866 \times mag_{\rm{IRAC}\,4} + 1.024.
\label{equ_iso}
\end{equation}
The slopes and zero points of this linear fit are similar for all fields, with a
cross-over ({\em Spitzer} IRAC\,4 equal to ISO 7\,$\mu$m magnitude) between
$6\fm0$ and $7\fm7$. The trend might be a reflection of a dependence of the
strength of the SiO fundamental band at $\sim 7.5$\,$\mu$m and/or a water band
at $\sim 5.5$\,$\mu$m on the brightness of the star. At least the SiO first
overtone band at $\sim 4.0$\,$\mu$m has been found to decrease in strength from
semi-regular variables to Mira-like variables \citep{Ari99}.
On top of that, instrumental effects in one or both missions cannot be
excluded.


\subsection{Comparison with GLIMPSE-II}\label{comp_GLIMPSE}

The Galactic Legacy Infrared Mid-Plane Survey Extraordinaire
\citep[GLIMPSE-II;][]{GLIMPSE}, is a {\em Spitzer} Legacy Science Programme of
most of the inner Galactic disk using the IRAC instrument. Our programme has
overlaps the GLIMPSE-II survey in the fields Bulge\,1, Bulge\,2, Bulge\,4, and
Bulge N\,1. A comparison with this large data set is especially interesting
because the same instrument and the same filter bands have been used, though
the observing strategies and data reduction techniques are different.

\begin{figure*}
\includegraphics{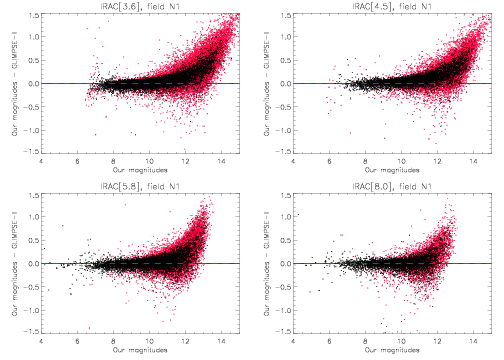}
\caption{Magnitude differences between our and the GLIMPSE-II catalogue for the
sources in common in the N\,1 field. Sources with quality flag~A in our
catalogue are represented by black dots, all other quality flags by red dots.}
\label{GLIMPSE_N1}
\end{figure*}

In Fig.~\ref{GLIMPSE_N1} we show the magnitude differences between our and
the GLIMPSE-II catalogue for all four IRAC channels for the Bulge N\,1 field,
since there is a large number of sources in common in this field (31\,191).
These plots look very similar for the other fields with overlap.
In general, the agreement between our and the GLIMPSE-II catalogue is good in
the bright to medium brightness range.
However, at the faint end, starting at around $10^{\rm{m}} - 11^{\rm{m}}$, we
notice a strong trend such that the magnitudes become increasingly fainter in
our catalogue than in the GLIMPSE-II catalogue. This is not only true for the
flag~B sources (most of the red dots at faint magnitudes in
Fig.~\ref{GLIMPSE_N1}), but also for flag~A sources fainter than this
magnitude limit, and for all filters. The same trend is revealed in a
comparison\footnote{http://www.astro.wisc.edu/sirtf/glm2\_galcen\_comparison.pdf}
between the GLIMPSE-II catalogue and the one of
\citet[][see next section]{Ram08}. According to B.\ Babler of the GLIMPSE-II
team (private communication, 2008), this trend at the faint magnitude limit
arises because of GLIMPSE-II's limitation to ``single frame'' photometry.
GLIMPSE-II does photometry only on single bcd frames instead of mosaics, because
each patch of the sky is observed only twice. This strategy may cause a
Malmquist bias, such that GLIMPSE-II magnitudes will be increasingly too bright
the closer a source is to the faint limit. Thus, for sources at the faint
limit, our magnitudes are likely more accurate than GLIMPSE-II magnitudes.
Finally, we also notice some saturation effects above $\sim6\fm5$ in IRAC\,1,
and a few sources that are significantly fainter in the GLIMPSE-II catalogue
than in our catalogue that can be found at all magnitudes.

\begin{figure*}[!ht]
\includegraphics[bb=66 364 549 720]{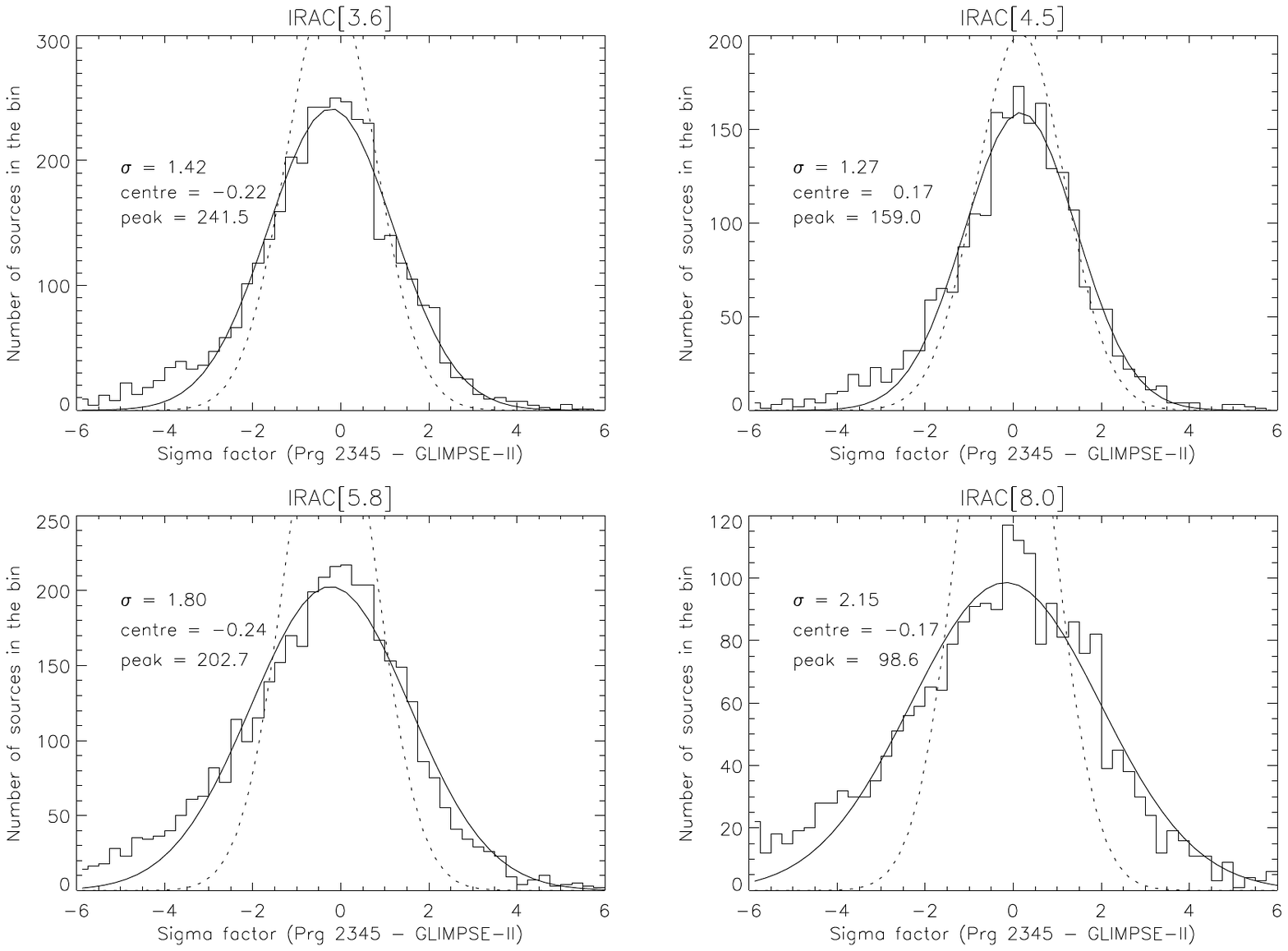}
\caption{Distribution of sigma-factors (Eq.~\ref{sigma}) for all IRAC channels
in the field N\,1 in the magnitude range $7\fm0 - 10\fm0$. The solid line is a
Gaussian fit to the distribution. The dotted line indicates a Gaussian of width
1 and the same area under the graph, the ideal case of correctly estimated
errors.}
\label{GLIMPSEerr_fig}
\end{figure*}

Due to the large number of sources in common with the GLIMPSE-II survey, we are
able to check whether or not the combined errors are of realistic magnitude. To
do so, we inspect the sigma-factor, which is also discussed in the
aforementioned comparison between the GLIMPSE-II and \citet{Ram08} catalogues:
\begin{equation}
\sigma = \frac{\rm{our~mag} - \rm{GLIMPSE~mag}}{\sqrt{(\rm{our~error})^2 + (\rm{GLIMPSE~error})^2}}.
\label{sigma}
\end{equation}
If the combined errors are close to the magnitude differences, the
sigma-factor will have a Gaussian distribution of width 1. If the errors are
underestimated, the distribution will be broader than that and narrower if the
error estimates are too large. Figure~\ref{GLIMPSEerr_fig} displays the sigma
distribution for all four IRAC channels for the field Bulge N\,1, along with a
Gaussian fit to the data. Also plotted is a Gaussian for a ``perfect''
distribution with the same area under the graph, i.e.\ when magnitude
differences and combined error are on average of the same size ($\sigma = 1$).
The source selection was restricted to the magnitude range $7\fm0 - 10\fm0$ to
avoid the tail of faint sources with a Malmquist bias (see above). We see that
the distributions are centred on zero, but there is a surplus of negative sigma
values, because of the sources with negative magnitude difference, as noted
above. The actual distributions for IRAC\,1 and 2 are only slightly broader
than what would be expected if the errors were correctly estimated. This means
that the combined errors for these channels are only slightly underestimated.
For IRAC\,3 and 4, the distributions are definitely broader than in the ideal
case. For these channels we have to assume that the combined errors are
underestimated by factor of 1.8 to 2.2. Indeed, our error estimates are on
average smaller than the ones estimated by GLIMPSE-II, but this is mostly due
to the longer exposure time per pixel in our survey, and we are doing the
photometry on the mosaicked images. If we used two times the error as given by
GLIMPSE-II in Eq.~\ref{sigma}, the sigma distribution would have the perfect
width of 1.0 for IRAC\,1, would be slightly too narrow for IRAC\,2 (too large
errors), and would still be too broad for IRAC\,3 and 4 (too small errors). We
therefore conclude that our error estimates for IRAC\,1 and 2 are probably only
slightly too small, and too small by a somewhat larger amount for IRAC\,3 and
4. The GLIMPSE-II errors, on the other hand, are probably correctly estimated
for IRAC\,1, slightly too large for IRAC\,2, and slightly underestimated for
IRAC\,3 and 4. Table~\ref{GLIMPSEerr_tab} summarises the width, centre, and
peak values of the Gaussian fits to the sigma distributions of
Fig.~\ref{GLIMPSEerr_fig}.

\begin{table}
\caption{Width, centre, and peak values of the sigma distributions.}
\label{GLIMPSEerr_tab}
\centering
\begin{tabular}{crrr}
\hline\hline
Band & $\sigma$ & Centre & Peak \\
\hline
IRAC\,1  & 1.42 & $-0.22$ & 241.5 \\
IRAC\,2  & 1.27 &    0.17 & 159.0 \\
IRAC\,3  & 1.80 & $-0.24$ & 202.7 \\
IRAC\,4  & 2.15 & $-0.17$ &  98.6 \\
\hline
\end{tabular}
\end{table}

\subsection{Comparison with \citet{Ram08}}

The overlap with the GALCEN survey of \citet{Ram08} is limited to the Bulge
N\,1 field. The reason for this is that the {\em Spitzer} time allocation
committee tried to avoid redundant observations of the same fields of sky
within GO programmes (GLIMPSE-II, on the other hand, is a legacy survey). A
comparison between our own and the GALCEN observations is useful, because for
our N\,2 field we will have to use the IRAC observations of GALCEN.
Nevertheless, because IRAC channels~1 and 3 are seeing a patch of the sky
neighbouring that of channels~2 and 4, we got a small overlap with
\citet{Ram08} in our Bulge N\,1 field in order to fully sample that field in
all bands. Adopting a stringent 1\farcs0 search radius, there are 5172 sources
in common in any one of the four IRAC bands. In the Bulge N\,2 field, 337
sources detected with MIPS\,24 have a counterpart in \citet{Ram08}, adopting a
less stringent 3\farcs0 search radius because of the less precise MIPS
coordinates.

\begin{figure*}[!th]
\includegraphics{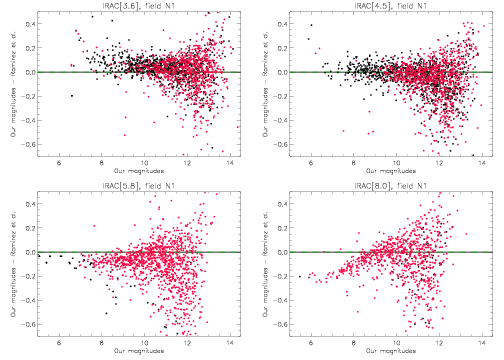}
\caption{Magnitude differences between our photometry and that of \citet{Ram08}
  for the small overlap in field N\,1. Symbols as in Fig.~\ref{GLIMPSE_N1}.}
\label{ramirez_fig}
\end{figure*}

Figure~\ref{ramirez_fig} shows the magnitude differences found for the sources
in common with \citet{Ram08} in the Bulge N\,1 field. The agreement in the
IRAC\,1 and 2 bands is very good. In the IRAC\,3 band, the GALCEN magnitudes
(of flag~A sources in the medium brightness range $7\fm0 - 10\fm0$) are on
average $\sim0\fm05$ fainter than our magnitudes. In the IRAC\,4 band, a
``knee'' appears for sources brighter than $8\fm2$. Since the same effect is
found in the aforementioned comparison between GLIMPSE-II and GALCEN data, the
origin of this ``knee'' most probably comes from an underestimation of fluxes
for sources brighter than $\sim8\fm2$ in IRAC\,4 by GALCEN. We will have to
consider these differences between our IRAC photometry and that of
\citet{Ram08} when comparing the results for Bulge N\,2 to those of other
fields.

The differences between our reduction and that of GALCEN, on the one hand, and
GLIMPSE-II, on the other, should be noted. While in our reduction the
background is determined for each pixel in a $45\times45$ pixel box centred on
that pixel and subtracted prior to the photometric measurements, GALCEN uses
the local background determined by PRF fitting and subtracts it from the
corresponding aperture flux of each detected source. For the fitting, PRFs
pre-defined by the SSC are used. GALCEN uses the aperture corrections as
recommended by the SSC, but adopts the PRF fluxes for their final catalogue
(except for sources where the ratio between PRF flux and aperture flux exceeds
1.5; for those sources the aperture flux is adopted). We adopted the PRF flux
in our catalogue, derived from the most recent mean PRFs provided by the SSC.
GLIMPSE-II, on the other hand, performs photometry on individual bcd frames,
not on mosaics. A combination of PRF fitting and aperture photometry called
``tweaking''\footnote{http://www.astro.wisc.edu/sirtf/glimpse\_photometry\_v1.0.pdf}
is used to measure the flux of a source. PRFs dynamically determined from each
individual frame are fitted to the sources and subtracted from the image. On
the residual image, aperture photometry was performed around the positions where
PRFs had been subtracted in the previous step. If the resultant aperture flux
is substantially positive or negative, the source has been under- or
over-subtracted. This residual aperture flux is then subtracted from or added
to the PRF flux to compensate for the under- or over-subtraction. However,
tweaking was applied rather sparingly and only in the IRAC\,1 and IRAC\,2 bands
of the GLIMPSE-II observations. These differences in reduction have to be kept
in mind, just as for the comparison between GALCEN and GLIMPSE-II.

\subsection{Comparison of MIPS\,24 measurements with \citet{Hin08}}

A catalogue of {\em Spitzer}/MIPS 24\,$\mu$m sources towards the region of
$\sim 1\fdg5 \times 8\fdg0$ around the Galactic centre has recently been
presented by \citet{Hin08}. With this survey, we have 371 sources in common
(i.e.\ unique counterparts within a one physical pixel = 2\farcs55 search
radius) in our Bulge\,4 field, 280 in the Bulge N\,1 field, and 129 in the
Bulge N\,2 field. Thanks to the longer exposure time of our observations
(331\,s per pixel compared to 15\,s per pixel in the fast scan mode used by
Hinz et al.), our catalogue reaches much fainter flux levels.

\begin{figure}
\includegraphics[width=\columnwidth,bb=69 368 546 703]{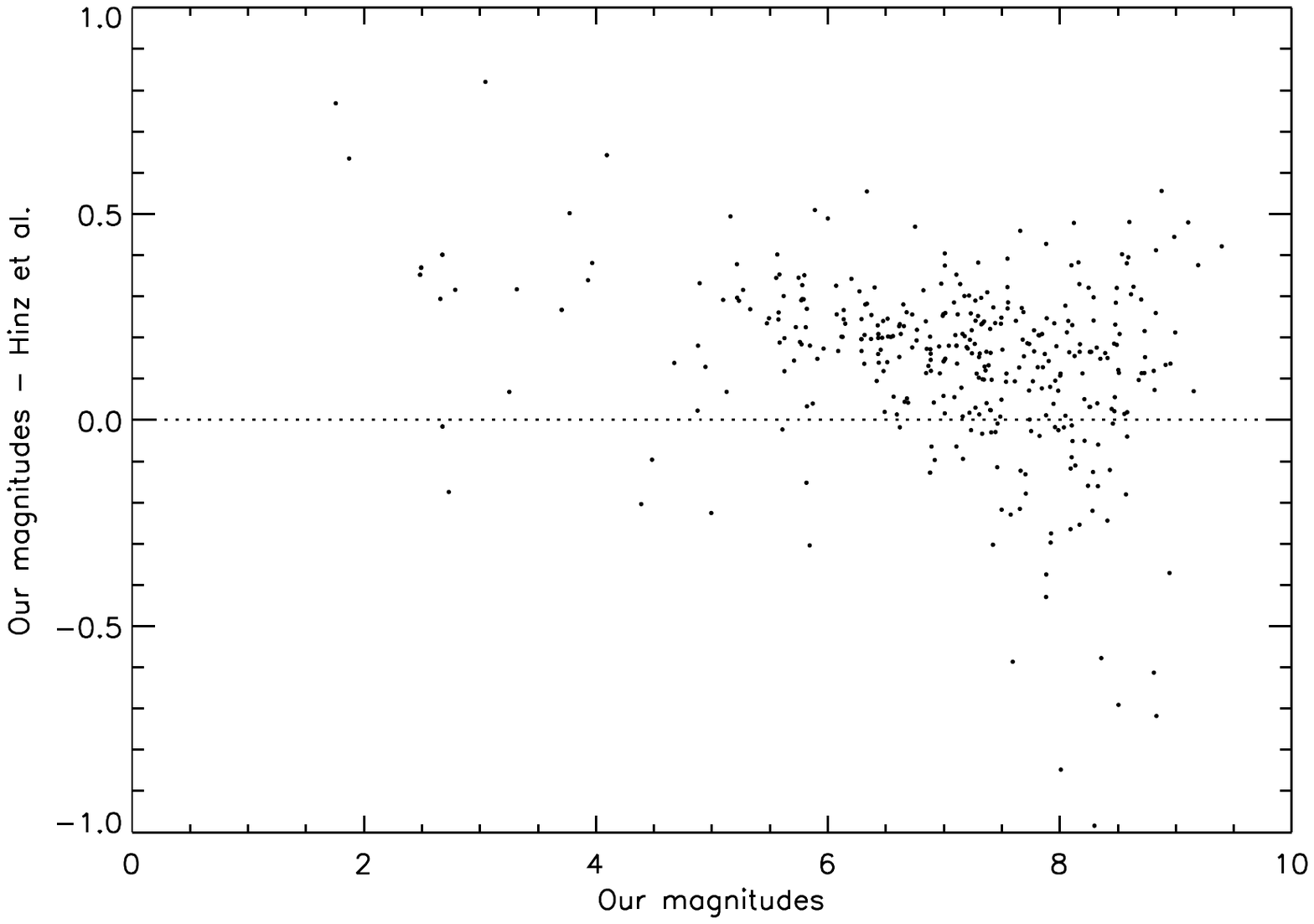}
\caption{Magnitude differences between our MIPS\,24 magnitudes and
\citet{Hin08}, for our Bulge\,4 field.}
\label{comp_MIPS_fig}
\end{figure}

For the N\,1 field, we only have a small overlap with the survey of
\citet{Hin08}. In the N\,2 field there is strong diffuse galactic emission,
and a small fraction of the sources might actually be false detections. We thus
concentrate on the Bulge\,4 field for a comparison with the \citet{Hin08}
catalogue. This is also the field with the most sources in common. In
Fig.~\ref{comp_MIPS_fig} we show the magnitude differences between our
measurements and that of \citet{Hin08} as a function of the source magnitude in
our catalogue. There might be a general offset of $\sim0\fm15$ in the sense
that our magnitudes are on average fainter, but the scatter is considerable. At
least there seems to be no trend at the faint magnitude end.

As for the comparison with GLIMPSE-II, we also checked the error estimates of
the MIPS\,24 data by investigating the distribution of sigma-factors
(Eq.~\ref{sigma}). The result of this exercise is that the combined errors are
too small by a factor of six or more. Even adopting twice the error estimate of
\citet{Hin08}, which is somewhat larger than ours, instead of a combination of
the errors, gives a sigma distribution that is wider by a factor of four than
what is expected from a correct estimate. We thus conclude that the errors are
still largely underestimated in both catalogues, and that the uncertainty due
to crowding and strong background radiations renders a precise flux
determination in the vicinity of the Galactic centre impossible. Some part of
the found differences, however, may be explained by a real variability, since
many of the bright sources are expected to be AGB variables in the bulge
\citep{Gla09}. At a wavelength of 20\,$\mu$m, a full amplitude of up to 0\fm7
is found for C-rich pulsating AGB variables in the solar neighbourhood
\citep[see e.g.\ Fig.~7 of][]{LeB92}.

\section{Colour-magnitude diagrams and comparison with theoretical isochrones}
\label{sec_cmds}

A final check on the quality of our data reduction are colour-magnitude diagrams
(CMDs) in combination with theoretical isochrones. If the flux measurement has
no systematic error (and also the isochrones are producing realistic colours),
the observed giant branch will be described well by the isochrone. This check is
restricted to the IRAC bands, since for many sources the MIPS\,24 band will be
affected by dust emission, which is very difficult to model accurately in
isochrones. We chose to use the most recent isochrones from \citet{Mar08}. A
web-form can be used for computing these isochrones with different
parameters\footnote{http://stev.oapd.inaf.it/cgi-bin/cmd\_2.1}. We computed an
isochrone with an age of 10\,Gyrs, me\-tal\-li\-ci\-ty $Z = 0.019$ (solar
me\-tal\-li\-ci\-ty), and no dust formation. In the mid-IR, the isochrones are
not very sensitive to the precise choice of the parameters such as age,
me\-tal\-li\-ci\-ty, and dust formation.

To disentangle possible systematic errors in the IRAC channels, we want to
include a flux measurement that is independent of our reduction method in the
comparison with the isochrone. Since all our fields are covered by 2MASS, which
can be regarded as a reliable source of $K_{\rm{S}}$ magnitudes, we decided
to construct dereddened $K_{\rm{S,0}}$ vs.\ $(K_{\rm{S}} - \rm{IRAC})_0$ CMDs.

Most of the fields towards the Galactic bulge suffer from strong extinction.
Thus, it is necessary to correct the photometric measurements for this
extinction before comparing them to the theoretical isochrone. Several
extinction maps for the Galactic centre region have been published.
\citet{Sch99} derived the extinction from the shift of the giant branch in the
$K_{\rm{S}}$ vs.\ $(J - K_{\rm{S}})$ CMD based on DENIS data. The extinction
values published in that work are given in the $V$ band. Another map of $A_{V}$
values towards fields in the Galactic bulge is presented by \citet{Sum04}.
These are based on the extinction measured on red clump giants in fields of the
OGLE-II survey \citep{Uda00}. Finally, the map of \citet{Dut03} uses the same
technique as \citet{Sch99} and applies it to 2MASS data. That map gives the
reddening value in the $K_{\rm{S}}$ band. Its spatial resolution is lower than
that of the other maps, but it is the only one that fully covers all of our
{\em Spitzer} fields.

\begin{table}
\caption{Coverage of our fields by different extinction maps of the Galactic
centre region.}
\label{tbl_maps}\centering
\begin{tabular}{lrr}
\hline\hline
Field & Ref.\ & median $A_{V}$\\
\hline
Bulge 1    & (1), 3  &  4.38 \\
Bulge 2    & 2, 3    &  2.45 \\
Bulge 3    & (1), 3  &  2.00 \\
Bulge 4    & 1, 3    & 15.75 \\
Bulge N\,1 & 1, 3    &  7.25 \\
Bulge N\,2 & 1, 3    & 20.06 \\
NGC\,6522  & 1, 2, 3 &  1.13 \\
\hline
\end{tabular}\\
\flushleft
References: 1: \citet{Sch99}; 2: \citet{Sum04}; 3: \citet{Dut03}.
Numbers in brackets mean that the field is not entirely covered by the
respective map, but to a significant fraction.
\end{table}

\begin{figure*}[!ht]
\includegraphics{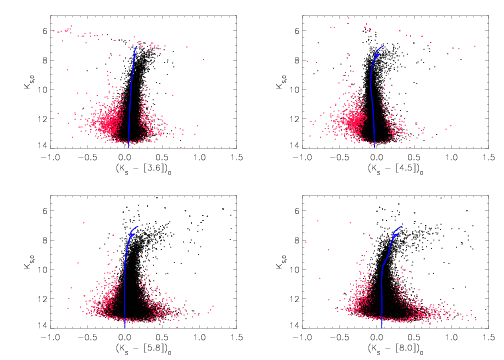}
\caption{Colour-magnitude diagrams of the field NGC\,6522 involving the 2MASS
$K_{\rm{S}}$ and different IRAC magnitudes. For this version, the extinction
map of \citet{Sch99} has been used. A theoretical isochrone from
\citet{Mar08} is included as a blue line (see text for details). Symbols are
as in Fig.~\ref{GLIMPSE_N1}.}
\label{NGC6522_CMD}
\end{figure*}

Table~\ref{tbl_maps} summarises the coverage of our seven fields by these
extinction maps, and the median extinction in the $V$ band is given. The
NGC\,6522 field is the only one that is covered by all three extinction maps.
Since this field also has the lowest extinction of all our {\em Spitzer} fields
so that uncertainties in the extinction determination should play a minor
role, it serves here as a ``fiducial'' field for the comparison with
theoretical isochrones in CMDs. We used the average values for the diffuse
inter-stellar medium determined by \citet{Ind05} for the extinction in the IRAC
bands:
$A_{K,\rm{S}} : A_{3.6} : A_{4.5} : A_{5.8} : A_{8.0} = 1 : 0.56 : 0.43 : 0.43 : 0.43$,
as well as $A_V : A_{K,\rm{S}} = 1 : 0.089$. Using the somewhat different
values found by other studies \citep[e.g.][]{Fla07,Nis09} has only a negligible
impact on the results. Figure~\ref{NGC6522_CMD} shows CMDs of the NGC\,6522
field involving the 2MASS $K_{\rm{S}}$ magnitude and the magnitudes of the four
IRAC bands.

The following observations can be made from these CMDs. At the faint end, the
distribution in colour is quite broad since the errors become quite large for
these faint sources, and some of the objects probably are background galaxies.
There are also a small number of objects with extremely red or extremely blue
colours. Possibly, false identifications are involved in some cases (we include
only sources with a 2MASS counterpart within 2\farcs0 search radius for the
comparison with CMDs). At intermediate brightnesses, the distribution in colour
is much narrower, and only very few outliers can be found. At the bright end,
the number of sources far from the isochrone locus increases again. This is
partly due to saturated sources, e.g.\ in IRAC\,1 (blue ``sequence'' at the
bright end), and to dust emission that already shows up at IRAC\,4 wavelengths.
In general we find good agreement between the observed and the predicted
location of the giant branch in the CMDs. In the
$(K_{\rm{S}} - \rm{IRAC\,1})_0$ CMD (upper left panel of
Fig.~\ref{NGC6522_CMD}), the tip of the giant branch in the isochrone is
somewhat bluer than the observed tip. In the $(K_{\rm{S}} - \rm{IRAC\,3})_0$
CMD (lower left panel), the theoretical isochrone is bluer than the observed
giant branch by about 0\fm1 over the whole brightness range. In this filter,
the agreement with the GLIMPSE-II survey is very good, but the GALCEN
magnitudes are fainter than our magnitudes. Thus, by adopting the GALCEN
magnitude scale, the shift of the isochrone would be somewhat alleviated, but
not all catalogues and the isochrone can be made agree. We do not find any
trends in magnitude in the IRAC\,1 filter in the comparison with any of the
other surveys, thus we assume that the difference is related to a small problem
with the isochrone in the $(K_{\rm{S}} - \rm{IRAC\,1})_0$ colour. The same
might be suspected for the $(K_{\rm{S}} - \rm{IRAC\,3})_0$ colour.

The comparison with the isochrones also allows for a check on how deep our
{\em Spitzer} photometry goes. We find that in all IRAC bands, we reach sources
down to the beginning of the He-core burning (horizontal branch), although
probably with a reduced detection probability and less accurate photometry. The
He-core burning phase is covered until its end, hence also the whole AGB
evolution. However, the sensitivity of the present observations is still far
too low to reach the beginning of the RGB phase, and our investigations will be
limited to the brighter half of the RGB. The MIPS photometry does not reach
down to the horizontal arm. The faintest sources that are detected with MIPS in
the NGC\,6522 field are roughly $3\fm5$ fainter than the RGB tip of the
isochrones.

As for the colour dependence on the used extinction map, we find that for
individual sources the de-reddened $K_{\rm{S}} - \rm{IRAC}$ colour may vary
by a few 0\fm01. However, no general trend is observed when different extinction
maps are applied. Thus, though the fine structure of the extinction varies
somewhat from map to map, the average magnitude of the extinction is very
similar among the maps.
%

\section{Summary and conclusions}\label{sec_conclu}

We present a catalogue of {\em Spitzer} IRAC/MIPS observations of seven fields
towards the Galactic bulge, sampling a range of galactocentric radii. These
observations allow us, amongst other things, to study the mass loss of a large
and homogeneous sample of RGB and AGB stars down to lower luminosities and
mass-loss rates than previously achieved.

In this first paper, we present the observations, the data reduction procedure,
and comparisons to other mid-IR surveys of the Galactic bulge. In general, we
find good agreement with other surveys. The comparison between the
{\em Spitzer} IRAC\,4 band and the ISOGAL LW2 band shows good agreement, but
reveals a slight trend with magnitude. GLIMPSE-II magnitudes are in good
agreement with our magnitudes in the bright-to-medium brightness range, but a
strong trend is present at the faint end. This trend is probably related to a
Malmquist bias in the GLIMPSE-II data set. The error estimates of GLIMPSE-II
and our IRAC photometry are reasonable in the IRAC\,1 and 2 bands, but somewhat
too small in the IRAC\,3 and 4 bands. In the comparison with GALCEN, we find on
average $\sim0\fm05$ brighter magnitudes in the IRAC\,3 band. The source of the
discrepancy at the bright end of the IRAC\,4 band is probably not related to
our catalogue. A comparison with the MIPS\,24 catalogue of \cite{{Hin08}}
reveals that our magnitudes are probably brighter at the level of $\sim0\fm15$,
although with larger scatter. We also find good agreement between our data
and recent isochrones in colour-magnitude diagrams for at least three of the
four IRAC bands. The $\sim0\fm1$ offset from isochrones involving IRAC\,3
deserves some more attention. We thus may assume that the observations and
data reduction are accurate on the level of $\sim0\fm1$ or better, as well as
precise on the level of $\lesssim0\fm1$, except for faint sources with quality
grade~B. The science exploitation of the data will follow in subsequent papers.

\begin{acknowledgements}
We thank Sean Carey for providing the artifact correction tools used for this 
work.

This research made use of Tiny Tim/{\em Spitzer}, developed by John Krist for
the {\em Spitzer} Science Center. The Center is managed by the California
Institute of Technology under a contract with NASA. 

The research described in this publication was partly carried out at the Jet
Propulsion Laboratory, California Institute of Technology, under a contract
with the National Aeronautics and Space Administration.

This work is based on observations made with the {\em Spitzer} Space Telescope,
which is operated by the Jet Propulsion Laboratory, California Institute of
Technology under a contract with NASA. Support for this work was provided by
NASA through an award issued by the JPL/Caltech.

This publication makes use of data products from the Two Micron All Sky Survey,
which is a joint project of the University of Massachusetts and the Infrared
Processing and Analysis Center/California Institute of Technology, funded by
NASA and the National Science Foundation.

SU acknowledges support from the Fund for Scientific Research of Flanders (FWO)
under grant number G.0470.07.

MSt has been supported by the European Community's Marie Curie Actions - Human
Resource and Mobility within the JETSET (Jet Simulations, Experiments and
Theory) network under contract MRTN-CT-2004 005592.

\end{acknowledgements}


\clearpage


\Online

\begin{appendix}
\section{De-blending or not de-blending?}
\label{sec_blending}

We tested the comparability of the fluxes derived by PRF fitting and those 
resulting from aperture photometry for the IRAC 3.6\,$\mu$m channel and  for
the field NGC\,6522 using the two de-blending capabilities in APEX. The first
method is passive de-blending, where already during the detection process
multiple peaks in a cluster of pixels are found and this information is used 
in the PRF fitting module to simultaneously fit all the peaks. The second 
method is active de-blending, i.e.\ the number of assumed sources is iteratively
increased during PRF fitting to decrease the final $\chi^2$. 

\begin{figure*}[!ht]
\includegraphics[height=0.45\textwidth,angle=-90]{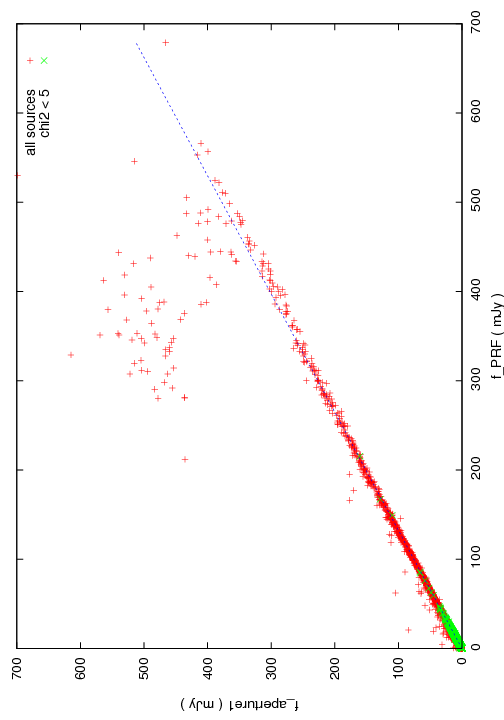}
\includegraphics[height=0.45\textwidth,angle=-90]{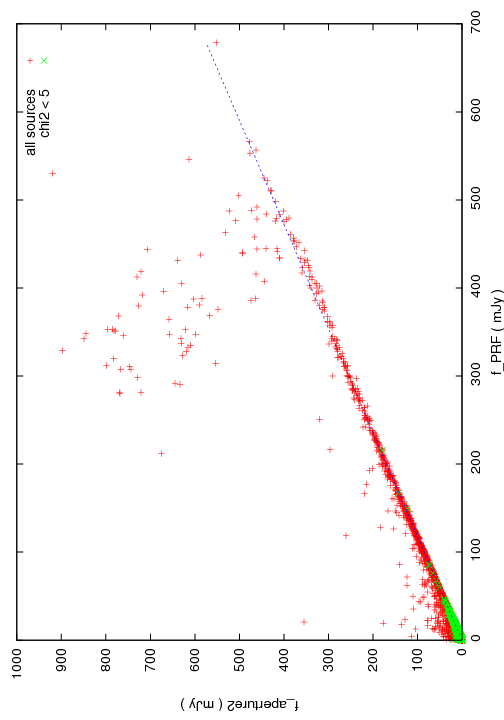}\\
\includegraphics[height=0.45\textwidth,angle=-90]{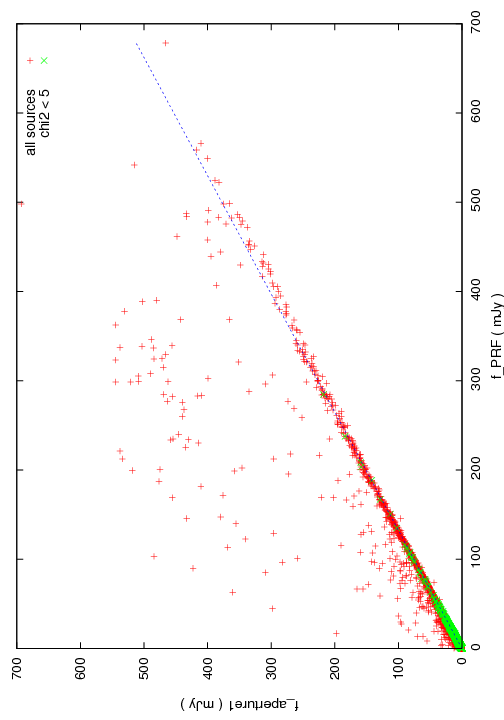}
\includegraphics[height=0.45\textwidth,angle=-90]{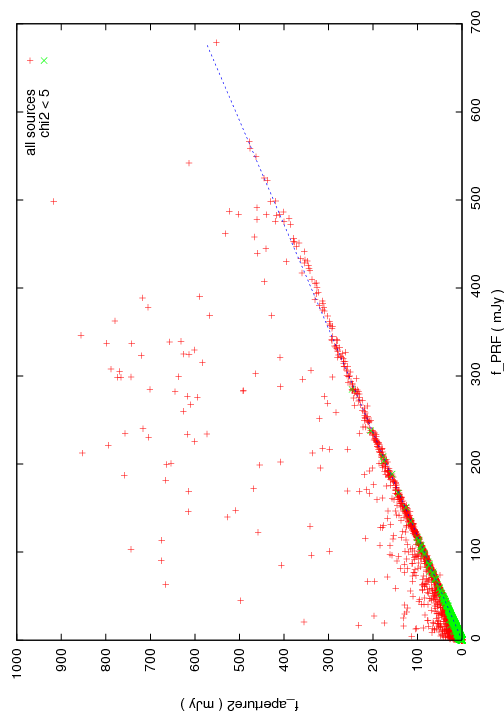}\\
\includegraphics[height=0.45\textwidth,angle=-90]{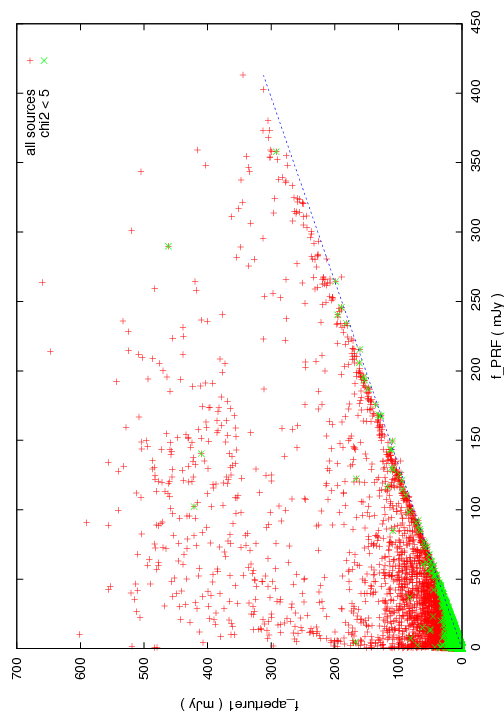}
\includegraphics[height=0.45\textwidth,angle=-90]{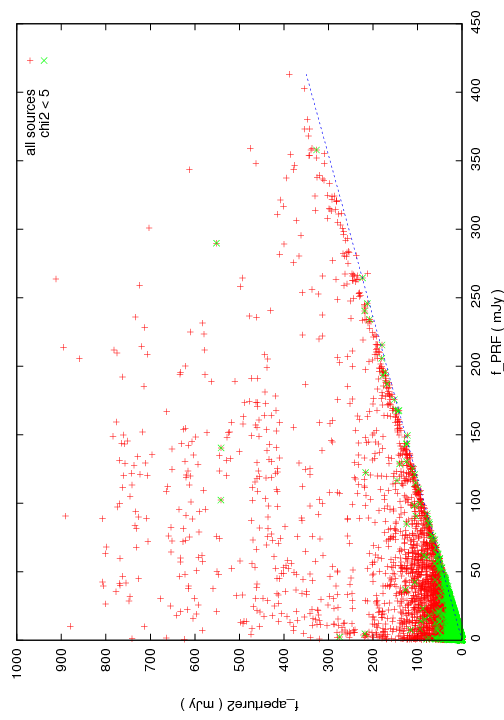}\\
\includegraphics[height=0.45\textwidth,angle=-90]{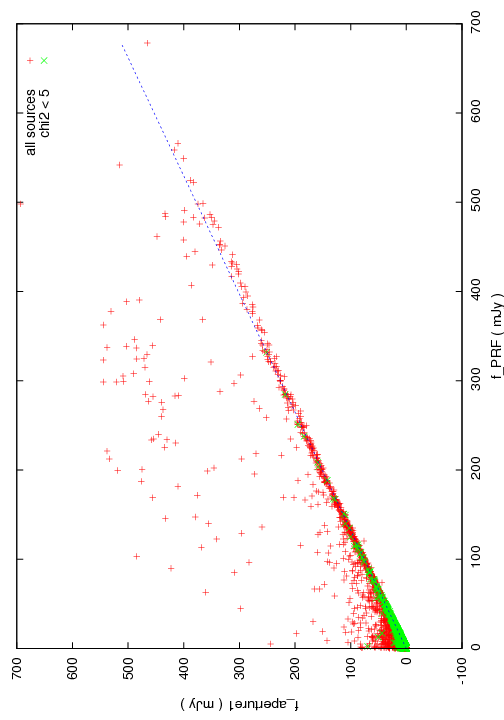}
\includegraphics[height=0.45\textwidth,angle=-90]{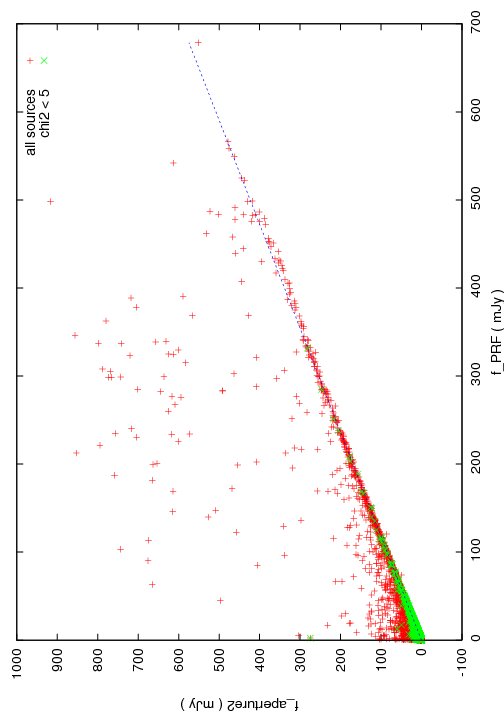}
\caption{Comparison of the fluxes resulting from PRF fitting and aperture 
photometry without de-blending (first row), with passive de-blending (second 
row), with active de-blending (third row), and with both (last row) for the 
apertures with radii of 2 pixels (left) and 3 pixels (right) for the field 
NGC\,6522 and IRAC\,1.}
\label{fluxes}
\end{figure*}

We use the point source probability images in the detection process and filter
the raw extraction table with the criterion that SNR is greater than 5 and that
APEX has not flagged the source as a point source outside the fitting area.
This excludes very weak sources, sources covered by only one image, and sources
whose position is not correctly identified. The numbers of detected and
extracted sources for the runs also shown in Fig.~\ref{fluxes} are given in
Table~\ref{tbl_extract_raw}, along with the number of sources filtered by the
two criteria.

\begin{table} 
\caption{Detected and extracted sources.}
\label{tbl_extract_raw}
\centering
\begin{tabular}{c|cc|ccc}
\hline\hline
        & detected & extracted & $SNR\leq5$ & sources outside & both criteria \\
\hline
no deblending          & 37974 & 37471 & 380 & 174 & 51 \\
pass. deblending       & 37974 & 37314 & 483 & 218 & 41 \\
act. deblending        & 70474 & 43655 & 2145 & 25885 & 1211 \\
act. \& pass. deblend. & 49679 & 39003 & 1743 & 9900 & 967 \\
\hline
\end{tabular}
\end{table}

The active de-blending procedure relies on the $\chi^2$ values to decide whether
de-blending is necessary or not, which results in heavy de-blending activity,
even for sources whose PRF and aperture fluxes are consistent. We therefore
disabled active de-blending in the final data reduction. 

Passive de-blending uses the information from the detection process about
whether different entries in the detection table belong to the same cluster of
pixels above the detection threshold. If yes, APEX fits the cluster
simultaneously, if not, the single sources one after the other. The first
effect of switching on passive de-blending is that the number of faint sources
that lie below an SNR value of 5, hence the number of sources rejected by the
first criterion, increases (see Table~\ref{tbl_extract_raw}). Since the
positions of the detected sources are also varied to fit the complete cluster,
the number of sources whose pointing is outside of the tile used for PRF fitting
increases, too (Table~\ref{tbl_extract_raw}). The number of extracted sources
is thus less than in the first run when passive de-blending is not activated.
Also brighter sources are affected, when there are other stars within 5--10
pixels. Then the independent stars belong to the same cluster and are fitted
simultaneously, although the PRF of one star does not contribute at all to the
flux in the small tile of another star used for PRF fitting. This again changes
the position of the centroid of the central bright star in the cluster and also
the flux measured by PRF fitting. Since the displacements are smaller than the
pixel size, the aperture flux does not change as much as the PRF flux. The
result is that the bright star moves away from the line of good agreement in
the f$_{\rm PRF}$-f$_{\rm aperture}$-plane (Fig.~\ref{fluxes}), so the scatter
around this line is increased. The same effect occurs for saturated and
super-saturated sources with a double peak leading to the observed spreading of
sources in the top-left part of Fig.~\ref{fluxes}. Thus the results of the run
with passive de-blending are not as reliable as the results of the run without
any de-blending.

\end{appendix}
\clearpage

\begin{appendix}
\section{Magnitude histograms of the seven bulge fields.}
\label{sec_histo}

\begin{figure*}
\includegraphics[width=\columnwidth,bb=60 363 552 720]{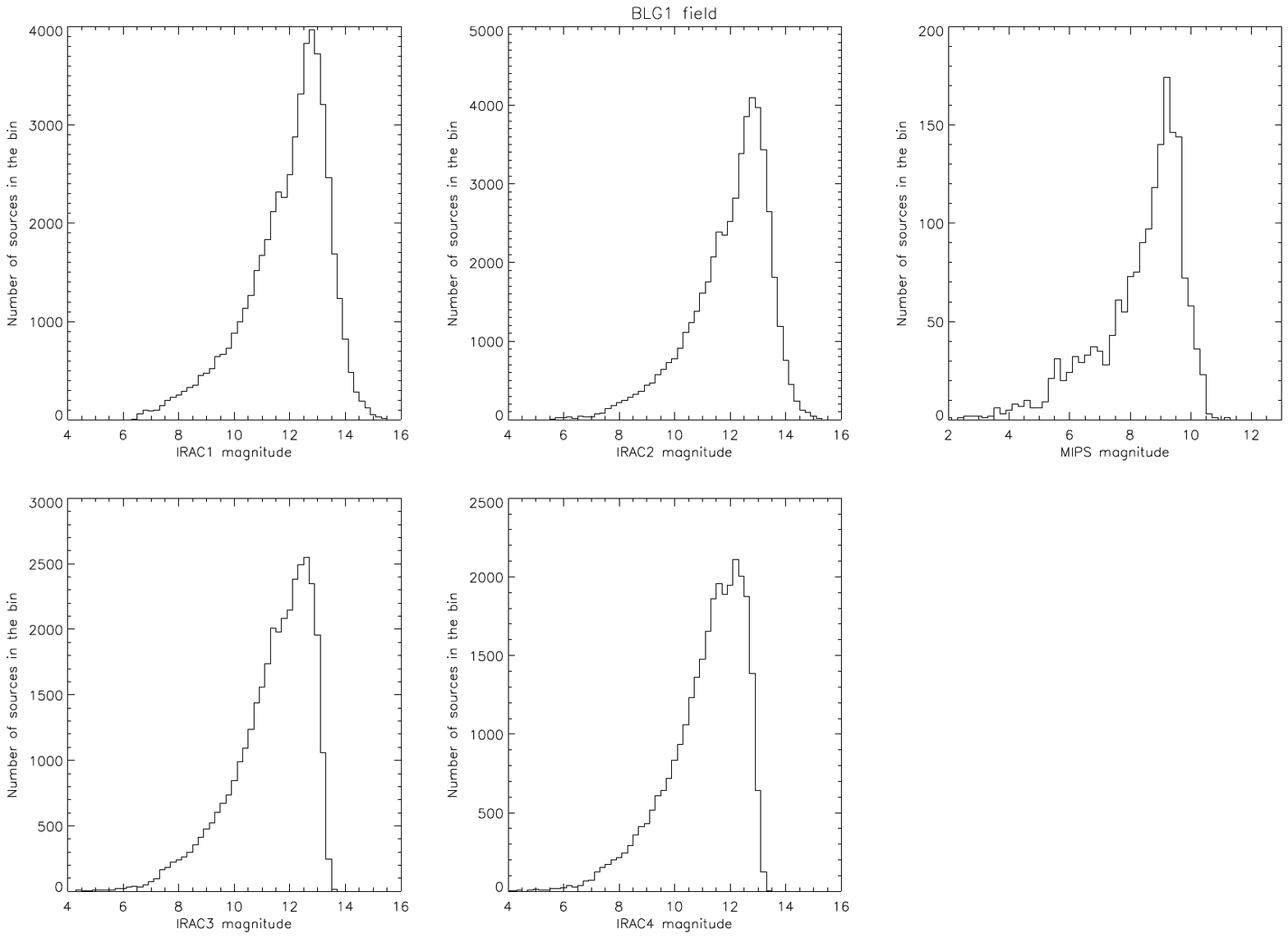}
\caption{Histogram of magnitudes for the Bulge\,1 field. The bin size is 0\fm2.}
\label{histblg1}
\end{figure*}

\begin{figure*}
\includegraphics[width=\columnwidth,bb=60 363 552 720]{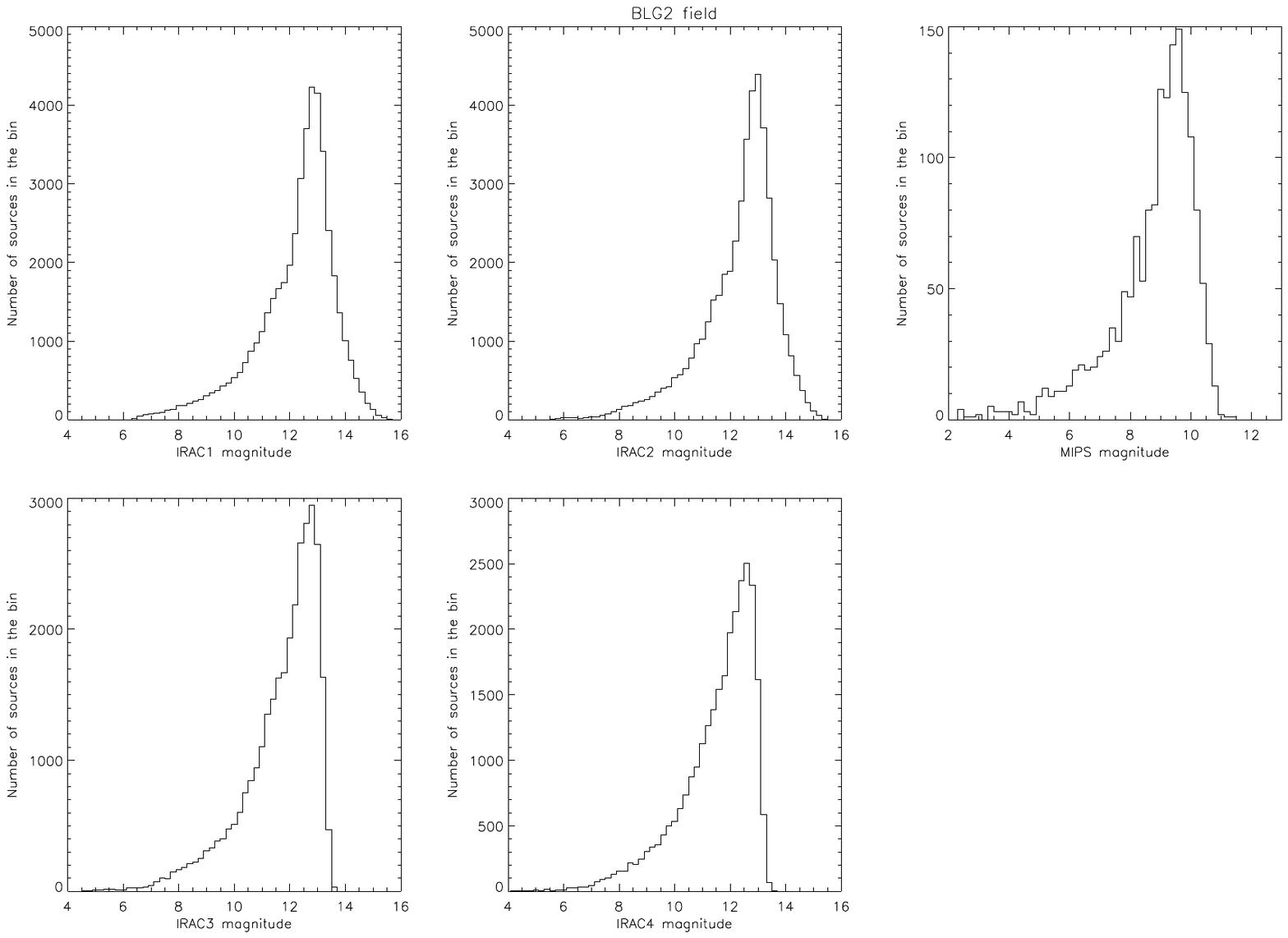}
\caption{Histogram of magnitudes for the Bulge\,2 field. The bin size is 0\fm2.}
\label{histblg2}
\end{figure*}

\begin{figure*}
\includegraphics[width=\columnwidth,bb=60 363 552 720]{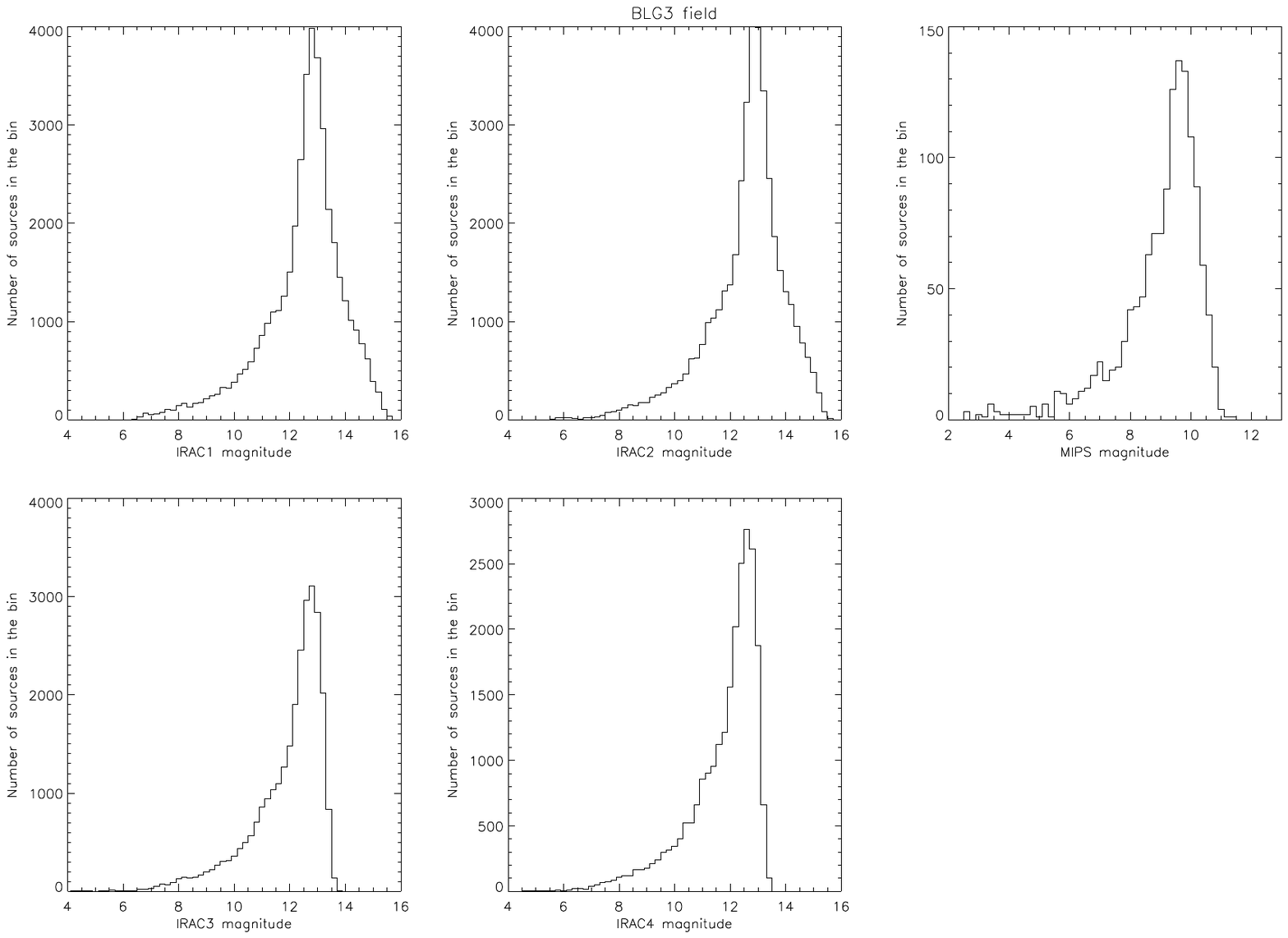}
\caption{Histogram of magnitudes for the Bulge\,3 field. The bin size is 0\fm2.}
\label{histblg3}
\end{figure*}

\begin{figure*}
\includegraphics[width=\columnwidth,bb=60 363 552 720]{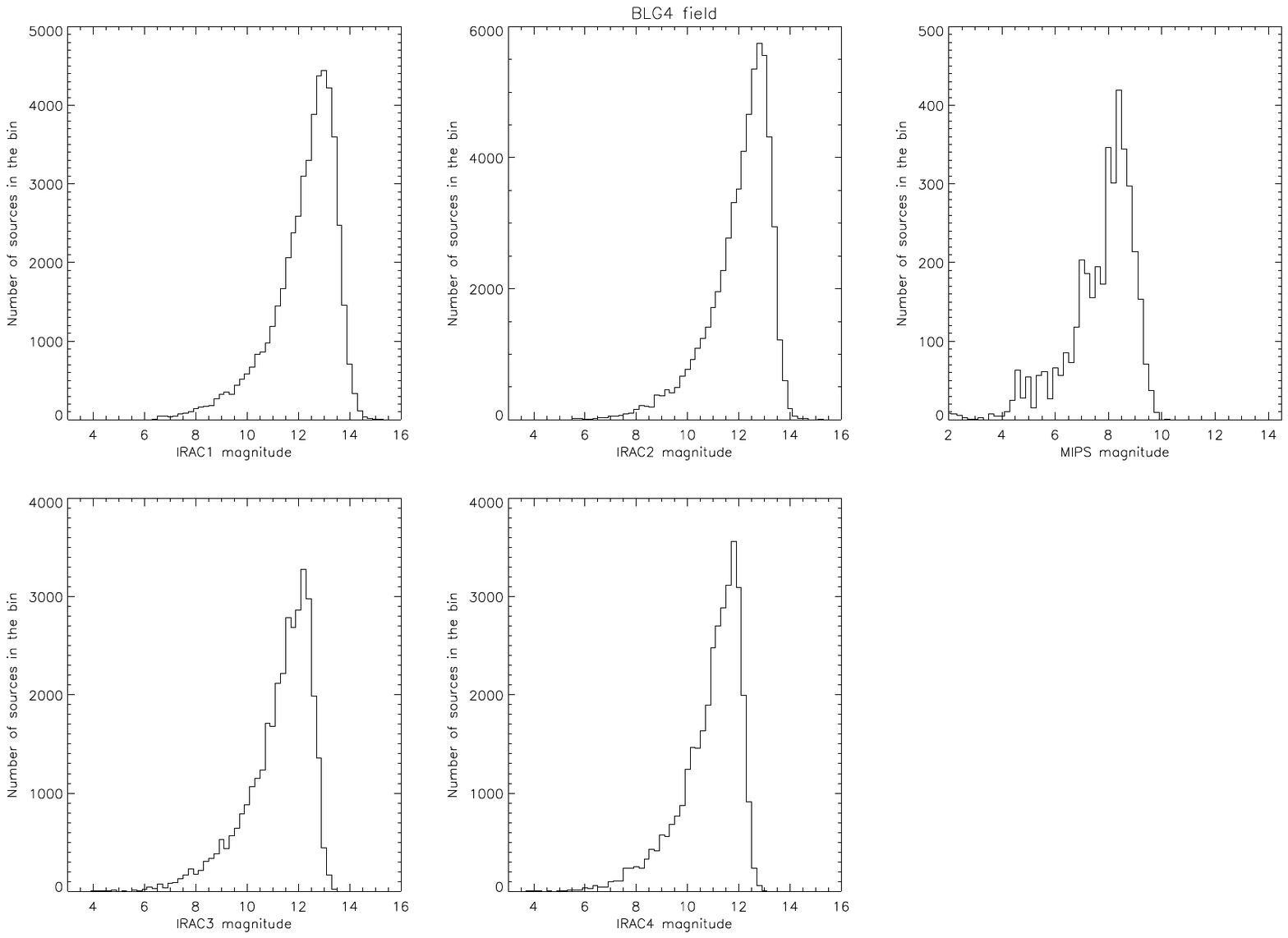}
\caption{Histogram of magnitudes for the Bulge\,4 field. The bin size is 0\fm2.}
\label{histblg4}
\end{figure*}

\begin{figure*}
\includegraphics[width=\columnwidth,bb=60 363 552 720]{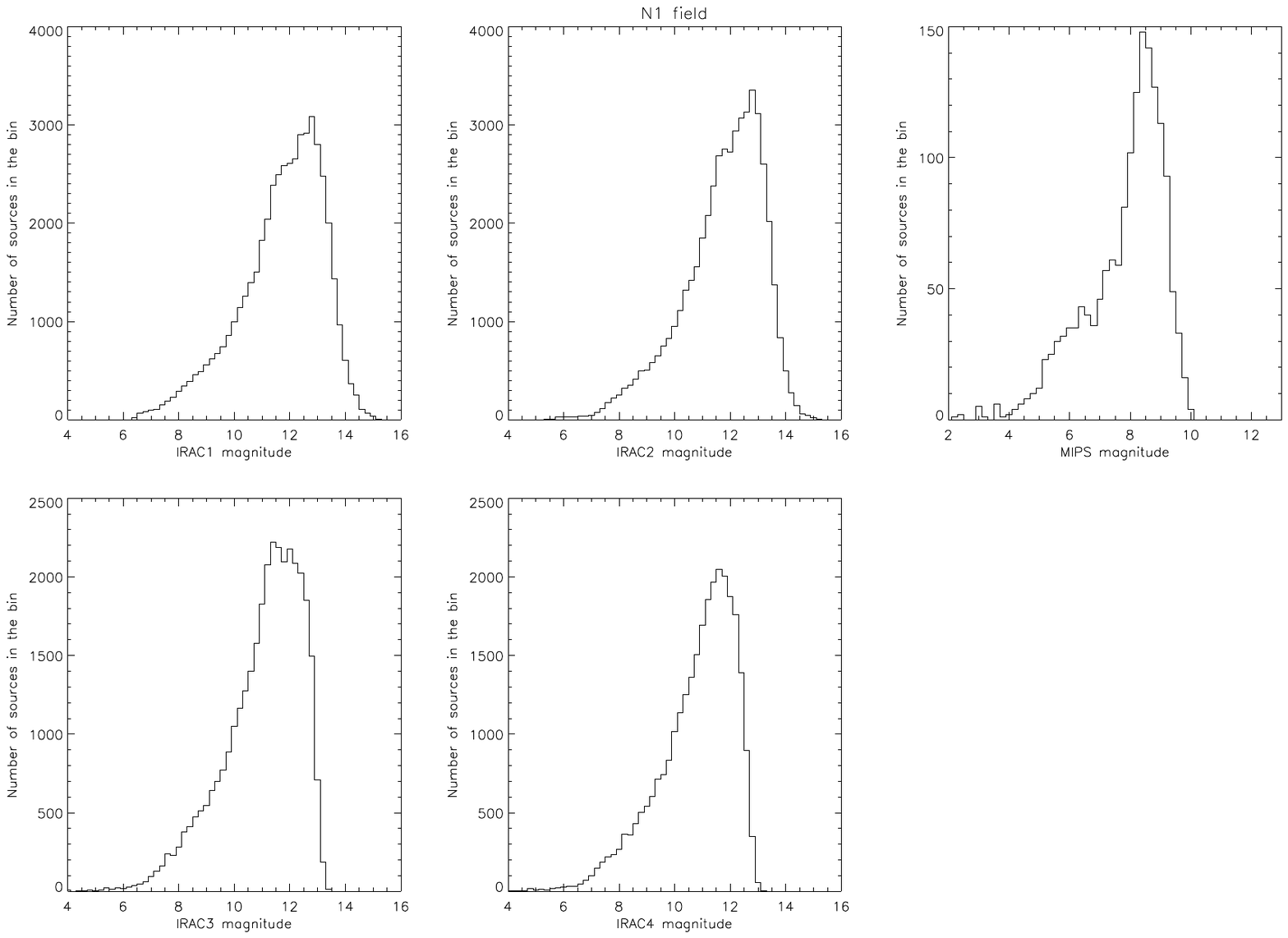}
\caption{Histogram of magnitudes for the Bulge N\,1 field. The bin size is 0\fm2.}
\label{histn1}
\end{figure*}

\begin{figure*}
\includegraphics[width=\columnwidth,bb=60 363 552 720]{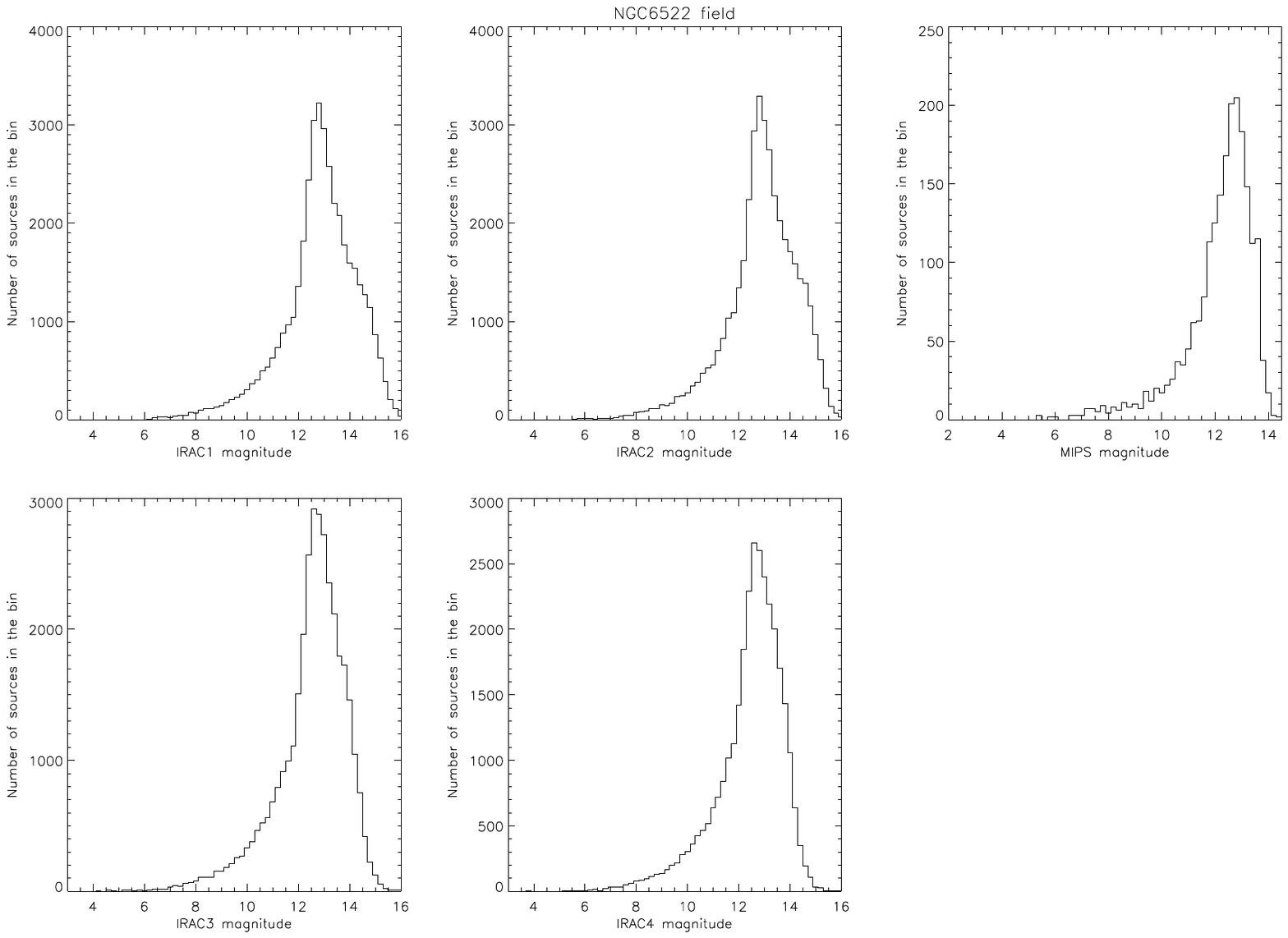}
\caption{Histogram of magnitudes for the NGC\,6522 field. The bin size is 0\fm2.}
\label{histngc6522}
\end{figure*}

\begin{figure*}
\includegraphics[width=\columnwidth,bb=60 363 552 720]{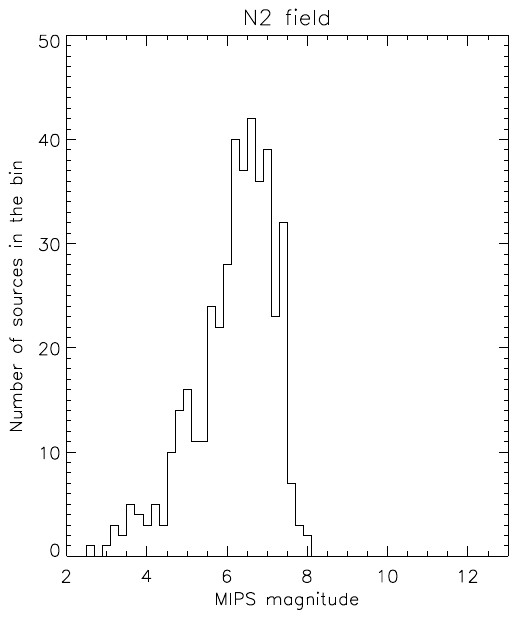}
\caption{Histogram of magnitudes for the Bulge N\,2 field. The bin size is 0\fm2.}
\label{histn2}
\end{figure*}

\end{appendix}

\end{document}